\documentclass{aa}

\usepackage{hyperref}
\usepackage{txfonts}
\usepackage{breakurl}

\newcommand {\mum} {$\muup$m}
\newcommand {\Hp} {$H_\mathrm{p}$}
\newcommand {\Ri} {$R_\mathrm{i}$}
\newcommand {\Rout} {$R_\mathrm{out}$}
\newcommand {\cmg} {cm$^2$g$^{-1}$}
\newcommand {\Mdisk} {$M_\mathrm{d}$}
\newcommand {\Msun} {$M_\odot$}

\begin{document}

\title{Brown dwarf disks with Herschel:\\Linking far-infrared and (sub)-mm fluxes\thanks{{\it Herschel} is a ESA space observatory with science instruments provided by European-led Principal Investigator consortia and with important participation from NASA.}}

\titlerunning{Brown dwarf disks}
\authorrunning{Daemgen et al.}

\author{Sebastian Daemgen\inst{\ref{inst1},\ref{inst2}}
\and Antonella Natta\inst{\ref{inst3},\ref{inst4}}
\and Alexander Scholz\inst{\ref{inst5}}
\and Leonardo Testi\inst{\ref{inst6},\ref{inst3},\ref{inst7}}
\and Ray Jayawardhana\inst{\ref{inst8}}
\and Jane Greaves\inst{\ref{inst9}}
\and Daniel Eastwood\inst{\ref{inst10}}}

\institute{ETH Z\"urich, Institut f\"ur Astronomie, Wolfgang-Pauli-Strasse 27, 8093 Z\"urich, Switzerland, \email{daemgens@phys.ethz.ch}\label{inst1}
\and Department of Astronomy \& Astrophysics, University of Toronto, 50 St.\ George Street, Toronto, ON, Canada M5H 3H4\label{inst2}
\and INAF/Osservatorio Astrofisico of Arcetri, Largo E. Fermi, 5, 50125, Firenze, Italy\label{inst3}
\and School of Cosmic Physics, Dublin Institute for Advanced Studies, 31 Fitzwilliams Place, 2, Dublin, Ireland\label{inst4}
\and SUPA School of Physics and Astronomy, University of St. Andrews, North Haugh, St. Andrews, Fife KY16 9SS, United Kingdom\label{inst5}
\and European Southern Observatory, Karl-Schwarzschildstr.\ 2, 85748, Garching, Germany\label{inst6}
\and Excellence Cluster ‘Universe’, Boltzmannstr. 2, D-85748 Garching bei M{\"u}nchen, Germany\label{inst7}
\and Faculty of Science, York University, 4700 Keele Street, Toronto, ON M3J 1P3, Canada\label{inst8}
\and School of Physics \& Astronomy, Cardiff University, Queen's Buildings, Cardiff CF24 3AA, UK\label{inst9}
\and Institute for Astronomy, University of Edinburgh, Royal Observatory, Blackford Hill, Edinburgh EH9 3HJ, U.K.\label{inst10}}

\abstract{
    Brown dwarf disks are excellent laboratories to test our understanding of disk physics in an extreme parameter regime.
    In this paper we investigate a sample of 29 well-characterized brown dwarfs and very low-mass stars, for which Herschel far-infrared fluxes and (sub)-mm fluxes are available. We measured new Herschel PACS fluxes for 11 objects and complement these with (sub)-mm data and Herschel fluxes from the literature. We analyze their spectral energy distributions in comparison with results from radiative transfer modeling.
    Fluxes in the far-infrared are strongly affected by the shape and temperature of the disk (and hence stellar luminosity), whereas the (sub)-mm fluxes mostly depend on disk mass. Nevertheless, there is a clear correlation between far-infrared and (sub)-mm fluxes. We argue that the link results from the combination of the stellar mass-luminosity relation and a scaling between disk mass and stellar mass. We find strong evidence of dust settling to the disk midplane. 
    The spectral slopes between near- and far-infrared are mostly between $-0.5$ and $-1.2$ in our sample, which is comparable to more massive T Tauri stars; this may imply that the disk shapes are similar as well, although highly flared disks are rare among brown dwarfs. We find that dust temperatures in the range of 7--15\,K, calculated with $T\approx25\,(L/L_\odot)^{0.25}$\,K, are appropriate for deriving disk masses from (sub)-mm fluxes for these low luminosity objects. About half of our sample hosts disks with at least one Jupiter mass, confirming that many brown dwarfs harbor sufficient material for the formation of Earth-mass planets in their midst.
    }


\maketitle

\section{Introduction}
Brown dwarfs (BDs) are a common outcome of star formation and have been found in large numbers in all nearby star-forming regions \citep[see review by][]{luh12}. Just like their more massive stellar siblings, young brown dwarfs are surrounded by dusty disks. Brown dwarfs have masses of 0.01--0.08$\,M_{\odot}$ and luminosities several orders of magnitude lower than stars, while their accretion rates and the disk masses are also lower. Their disks represent an interesting laboratory for studying the evolutionary processes in disks, particularly those related to planet formation, and our methods for inferring disk properties, in an extreme parameter range.

Brown dwarf disks, originally found in near- and mid-infrared surveys \citep{com98,mue01,nat01,nat02, jay03}, are now investigated over the infrared, submillimeter, and millimeter spectral range. For the most part, the general evolutionary blueprint adopted for stellar disks applies to brown dwarf disks as well. The disk lifetimes are 5--10\,Myr and thus comparable to or slightly longer than in the stellar regime \citep{daw13,luh12a}. At a given age, brown dwarf disks show a range of masses and geometries, including flat and flared disks as well as disks with inner opacity holes \citep{moh04}. As in stellar disks, evidence for the presence of large, mm-sized dust grains has been found by various methods \citep[e.g.,][]{apa05,sch07}, and was recently demonstrated based on the first brown dwarf data from the Atacama Large Millimeter Array (ALMA) \citep{ric12,ric14}.

Observations of brown dwarfs in the (sub)-mm and millimeter domain show that disk masses scale with stellar mass, at around 1\% of the stellar mass, albeit with large scatter \citep{sch06,kle03,moh13,and13}. Substellar disk masses rarely exceed 0.001$\,M_{\odot}$, thus only the brightest BD disks have been detected at these wavelengths. 

Only recently has it been possible to investigate brown dwarfs in the far-infrared between 70 and 160\,\mum, thanks to the Herschel space observatory. \citet{har12b,har12} carried out a Herschel survey of disks around $\sim 40$ very low-mass objects (VLMOs) at 70 and 160\,\mum. Comparing their data with radiative transfer models, they infer a wide range of disk masses (from $<$$10^{-6}$ to $10^{-3}\,M_{\odot}$). While the upper limit agrees with the values inferred from (sub)-mm observations, the minimum values are lower than expected. A few other groups have recently published their analyses of VLMO disks based on Herschel/PACS data (i.e., wavelength of 70--160\,\mum): \citet[detections for 12 BDs in $\rho$-Oph]{alv13}, \citet[detections for a few VLM stars in Chamaeleon]{olo13}, \citet[detections for 58 VLMOs in Taurus]{bul14}, and \citet[detections for 5 VLMOs in the TW Hydrae association]{liu15a}, most without (sub)-mm detection. One finding of these studies is that the geometrical parameters of the disk, in particular the flaring angle, seem to be similar in VLMOs compared with more massive stars. Finally, \citet{joe13} report a highly flared disk with a mass of $10^{-4}\,M_{\odot}$ for the isolated planetary-mass object OTS44, using a spectral energy distribution (SED) with a detection at 70\,\mum\ and an upper limit at 160\,\mum, but without (sub)-mm data points.

So far, the focus has been on the analysis of VLMO disks in specific wavelength domains (e.g., near-infrared (NIR), far-infrared (FIR), or (sub)-mm), but there is little work on the links between the different parts of the SEDs of these sources. In this paper, we set out to study the FIR and sub-mm to mm fluxes for a sample of well-characterized VLMOs with masses below 0.2\,\Msun, for which information is available in both these wavelength domains. This approach allows us to  investigate specifically physical properties of the disk that affect the long-wavelength portions of the SED, including the degree of flaring, disk mass, dust temperature, and the interdependence among these parameters.

\section{Sample and data}
\subsection{Observations and literature data}\label{sec:obs}
The present study combines our own Herschel observations of VLMOs with previously published Herschel data. We aim to select all VLMOs (defined as spectral type M4 or later) with well-sampled disk SEDs and well-known stellar parameters. More specifically, we focus on objects with a detection in at least one Herschel band and a detection at (sub)-mm wavelengths.

\subsubsection{New Herschel observations}

\begin{table*}
\caption{Stellar properties\label{tab:stellar}}
\centering
\tiny
\begin{tabular}{lcr@{\,}c@{\,}lr@{\,}c@{\,}lcccccccc}
\hline\hline
&
&
\multicolumn{3}{c}{RA} &
\multicolumn{3}{c}{DEC} &
&
$T_\mathrm{eff}$&
$J$\tablefootmark{a}&
$A_J$&
dist &
$L_\mathrm{bol}$&
$M_*$&
\\
Target&
2MASS ident  &
\multicolumn{3}{c}{(J2000)}  &
\multicolumn{3}{c}{(J2000)} &
SpT&
(K)&
(mag)&
(mag)&
(pc) &
($10^{-3}L_\odot$)&
($M_\odot$)&
ref\\
\hline
\multicolumn{15}{c}{\emph{Our Sample}}               \\
\hline
2M04152746                           &  J04155799$+$2746175  &  04&15&57.994  &  $+$27&46&17.57  &  M5.5   &  3054    &  11.745  &  0.16  & 140   &   60  & 0.14  & 1,5 \\
2M04232801                           &  J04230607$+$2801194  &  04&23&06.07   &  $+$28&01&19.5   &  M6     &  2990    &  12.242  &  0.21  & 140   &   39  & 0.096 & 2   \\
2M04332615                           &  J04334465$+$2615005  &  04&33&44.652  &  $+$26&15&00.53  &  M4.75  &  3161    &  11.639  &  0.85  & 140   &  125  & 0.20  & 1   \\
2M04412534                           &  J04414825$+$2534304  &  04&41&48.250  &  $+$25&34&30.50  &  M7.75  &  2763    &  13.730  &  0.28  & 140   &   10  & 0.036 & 2,5 \\
CFHT-6                               &  J04390396$+$2544264  &  04&39&03.960  &  $+$25&44&26.42  &  M7.25  &  2846    &  12.646  &  0.11  & 140   &   23  & 0.050 & 2,5 \\
CFHT-18                              &  J04292165$+$2701259  &  04&29&21.653  &  $+$27&01&25.95  &  M5.25  &  3089    &  10.801  &  0\tablefootmark{b}& 140 &  130  & 0.16  & 2 \\
CIDA-1                               &  J04141760$+$2806096  &  04&14&17.609  &  $+$28&06&09.70  &  M5.5   &  3054    &  11.726  &  0.79  & 140   &  106  & 0.14  & 2,5 \\
GM Tau                               &  J04382134$+$2609137  &  04&38&21.340  &  $+$26&09&13.74  &  M6.5   &  2935    &  12.804  &  0.037 & 140   &   19  & 0.075 & 2   \\
ISO-Oph 32                           &  J16262189$-$2444397  &  16&26&21.899  &  $-$24&44&39.76  &  M8     &  2710    &  12.340  &  0.00  & 130   &   23  & 0.030 & 3   \\
ISO-Oph 102                          &  J16270659$-$2441488  &  16&27&06.596  &  $-$24&41&48.84  &  M5.5   &  3054    &  12.433  &  0.13  & 130   &   27  & 0.14  & 3   \\
ISO-Oph 160                          &  J16273742$-$2417548  &  16&27&37.422  &  $-$24&17&54.87  &  M6     &  2990    &  14.148  &  1.70  & 130   &   21  & 0.096 & 4   \\
\hline                                                                                                                    
\multicolumn{15}{c}{\emph{Literature Sample}}          \\                                                                 
\hline                                                                                                                    
L1521F-IRS\tablefootmark{c}          &                       &  04&28&38.95   &  $+$26&51&35.1   & M6--M8  &$\sim$2880&  \dots   &  \dots &  140  &\dots  & 0.058 & 6,8 \\
FW\,Tau\,A+B+C                       &  J04292971$+$2616532  &  04&29&29.710  &  $+$26&16&53.21  &  M5.5   &  3054    &  10.340  &  0.00  &  140  &  190  & 0.14  & 6,8 \\
FN\,Tau                              &  J04141458$+$2827580  &  04&14&14.590  &  $+$28&27&58.06  &  M5     &  3125    &   9.469  &  0.00  &  140  &  432  & 0.18  & 6,8 \\
CFHT\,12                             &  J04330945$+$2246487  &  04&33&09.457  &  $+$22&46&48.70  &  M6     &  2990    &  13.154  &  0.45  &  140  &   21  & 0.096 & 6,8 \\
ZZ\,Tau\,IRS\tablefootmark{c}        &  J04305171$+$2441475  &  04&30&51.714  &  $+$24&41&47.51  &  M5     &  3125    &  12.842  &  0.82  &  140  &   40  & 0.18  & 6,8 \\
J04381486+2611399\tablefootmark{c}   &  J04381486$+$2611399  &  04&38&14.861  &  $+$26&11&39.94  &  M7.25  &  2846    &  15.176  &  0.26  &  140  &    2.6& 0.050 & 6,8 \\
CIDA\,7                              &  J04422101$+$2520343  &  04&42&21.017  &  $+$25&20&34.38  &  M4.75  &  3161    &  11.397  &  0.32  &  140  &   98  & 0.20  & 6,8 \\
KPNO-10                              &  J04174955$+$2813318  &  04&17&49.554  &  $+$28&13&31.85  &  M5     &  3125    &  11.889  &  0.34  &  140  &   63  & 0.18  & 6,8 \\
IRAS\,04158+2805\tablefootmark{c}    &  J04185813$+$2812234  &  04&18&58.138  &  $+$28&12&23.49  &  M5.25  &  3089    &  13.778  &  0.66  &  140  &   14  & 0.16  & 6,8 \\
J04202555+2700355                    &  J04202555$+$2700355  &  04&20&25.554  &  $+$27&00&35.55  &  M5.25  &  3089    &  12.861  &  0.42  &  140  &   27  & 0.16  & 6,8 \\
KPNO-3                               &  J04262939$+$2624137  &  04&26&29.392  &  $+$26&24&13.79  &  M6     &  2990    &  13.323  &  0.42  &  140  &   17  & 0.096 & 6,8 \\
MHO-6                                &  J04322210$+$1827426  &  04&32&22.109  &  $+$18&27&42.64  &  M4.75  &  3161    &  11.711  &  0.37  &  140  &   77  & 0.20  & 6,8 \\
CFHT\,4                              &  J04394748$+$2601407  &  04&39&47.484  &  $+$26&01&40.78  &  M7     &  2880    &  12.168  &  0.00  &  140  &   33  & 0.058 & 6,8 \\
ESO-HA\,559                          &  J11062554$-$7633418  &  11&06&25.549  &  $-$76&33&41.87  &  M5.25  &  3089    &  13.009  &  1.01  &  160  &   52  & 0.16  & 7   \\
2M0444+2512                          &  J04442713$+$2512164  &  04&44&27.132  &  $+$25&12&16.41  &  M7.25  &  2846    &  12.195  &  0.026 &  140  &   33  & 0.050 & 6,8 \\
TWA\,30B\tablefootmark{d}            &  J11062554$-$7633418  &  11&32&18.223  &  $-$30&18&31.65  &  M5     &  3125    &  15.350  &  0.00  &   46  &  0.2  & 0.12  & 9   \\
TWA\,32                              &  J12265135$-$3316124  &  12&26&51.365  &  $-$33&16&12.55  &  M6     &  2990    &  10.691  &  0.00  &   59  &   23  & 0.072 & 9   \\
TWA\,34                              &  J12265135$-$3316124  &  10&28&45.808  &  $-$28&30&37.46  &  M5     &  3125    &  10.953  &  0.00  &   47  &   12  & 0.12  & 9   \\
\hline
\end{tabular}
\tablefoot{
\tablefoottext{a}{$J$-band magnitude from 2MASS \citep{cut03,skr06}.}
\tablefoottext{b}{Zero extinction is assumed owing to the lack of better estimates. Previous measurements, e.g., $A_J=2$\,mag by \citet{luh09} have likely been corrupted by its binarity.}
\tablefoottext{c}{These targets were classified as likely Class\,I in Sect.~\ref{sec:phot} and are excluded from all disk analysis.}
\tablefoottext{d}{TWA\,30B likely has an edge-on disk \citep{loo10}.}
\tablebib{(1) \citealt{luh09}; (2) \citealt{fur11}; (3) \citealt{alv13}; (4) \citealt{gee11}; (5) \citealt{her14}; (6) \citealt{luh10}; (7) \citealt{luh07}; (8) \citealt{bul14}; (9) \citealt{liu15a}.}
}
\end{table*}
Observations for our program OT1\_ascholz\_1 (PI: A.~Scholz) were carried out on February 22 and March 12, 2013, in the three PACS \citep{pog10} bands at wavelengths of 70, 100, and 160\,\mum. {Targets were selected to be spectroscopically confirmed low-mass and substellar objects that show strong excess at mid-infrared wavelengths indicating the presence of an inner disk, are bright in the near- to mid-infrared ($F$(24\,\mum)\,$>$\,15\,mJy), have been extensively characterized at optical, NIR, and sub-mm wavelengths, and are well-isolated ($>$\,1$^\prime$) from nearest neighbors and strong cloud background at 24\,\mum. Out of 16 targets selected according to these criteria, 11 were observed, spanning spectral types from M4.75 to M8.} Table~\ref{tab:stellar} lists the most important parameters for the central objects and references.
Luminosities were inferred from the $J$-band magnitude and bolometric corrections in \citet{pec13}. Through a comparison with other bolometric correction tables \citep[e.g.,][]{har94}, we infer a systematic uncertainty on the order of $<$20\%.
Stellar masses were derived from effective temperatures, inferred from spectral types using Table~8 in \citet{luh03}, and BCAH98 isochrones \citep{bar98} assuming an age of 2\,Myr {based on their membership in the Taurus or $\rho$-Ophiuchus star-forming regions \citep{dae15,wil05}}. Inferred masses range from 0.03\,M$_\odot$ to 0.2\,M$_\odot$.

\subsubsection{Literature data}
Through an extensive literature search, we found 17 additional objects in the same mass range that have Herschel and (sub)-mm data points. Herschel data for these objects have been published by \citet{bul14}, \citet{har12}, \citet{how13}, \citet{olo13}, and \citet{kea14}. The (sub)-mm data come from a variety of sources. One more object, 2M0444, which ranks among the brightest disks in the VLM domain, has a reliable 70\,\mum\ flux from Spitzer/MIPS plus (sub)-mm data, and was also added to the sample. We derive stellar properties in the same way as for our core sample with the exception of the TWA Hydrae sources for which we assume an age of 10\,Myr {\citep{bel15}} to derive their masses. We list all derived values in Table~\ref{tab:stellar}. Flux measurements of the core and literature samples are presented in Table~\ref{tab:fluxes}.
\begin{table*}
\caption{Far-IR and (sub)-mm fluxes or 3$\sigma$ upper limits\label{tab:fluxes}}
\centering
\small
\begin{tabular}{lr@{\,$\pm$\,}lr@{\,$\pm$\,}lr@{\,$\pm$\,}lcr@{\,$\pm$\,}lr@{\,$\pm$\,}lr@{\,$\pm$\,}lc}
\hline\hline
{Target}        &
\multicolumn{2}{c}{F$_{70}$} &
\multicolumn{2}{c}{F$_{100}$} &
\multicolumn{2}{c}{F$_{160}$} &
{} &
\multicolumn{2}{c}{F$_{850}$} &
\multicolumn{2}{c}{F$_{890}$} &
\multicolumn{2}{c}{F$_{1300}$} &
{} \\
&
\multicolumn{2}{c}{(mJy)} &
\multicolumn{2}{c}{(mJy)} &
\multicolumn{2}{c}{(mJy)} &
{Ref} &
\multicolumn{2}{c}{(mJy)} &
\multicolumn{2}{c}{(mJy)} &
\multicolumn{2}{c}{(mJy)} &
{Ref}\\
\hline
\multicolumn{15}{c}{\emph{Our Sample}}               \\                                          
\hline              
2M04152746            &   25          &  3       &       26.8 &  3.0       &\multicolumn{2}{c}{$<$23.7} & 1,T,T &\multicolumn{2}{c}{\dots}& 32.9          &  15.2     & 12.6   &   1.4                & 4       \\
2M04232801            &   41          &  3       &       55.9 &  2.7       &  67.1 &    9.5             & 1,T,T &\multicolumn{2}{c}{\dots}& 13.3          &   6.6     &  5.1   &   1.1                & 4       \\
2M04332615            &  149          &  2       &      174.4 &  3.7       & 181.1 &   15.9             & 1,T,T &\multicolumn{2}{c}{\dots}& 47.8          &  21.9     & 18.3   &   1.7                & 4       \\
2M04412534            &   37          &  3       &       46.0 &  2.7       &  52.7 &   13.6             & 1,T,T &\multicolumn{2}{c}{\dots}&  5.7          &   2.8     &  2.2   &   0.4                & 4       \\
CFHT-6                &   23          &  3       &       27.4 &  2.4       &  43.6 &   13.4             & 1,T,T &\multicolumn{2}{c}{\dots}&  7.6          &   4.0     &  2.9   &   0.8                & 4       \\
CFHT-18               &  329          &  3       &      295.5 &  3.0       & 199.3 &   16.9             & 1,T,T &\multicolumn{2}{c}{\dots}&\multicolumn{2}{c}{\dots}  &\multicolumn{2}{c}{$<$3.5}     & 4       \\
CIDA-1                &  266          &  2       &      294.6 &  2.8       & 249.3 &   13.6             & 1,T,T &\multicolumn{2}{c}{\dots}& 27.0          &   0.3     & 13.5   &   2.8                & 5,6     \\
GM\,Tau               &   36          &  2       &       35.7 &  2.3       &  30.6 &    6.3             & 1,T,T &\multicolumn{2}{c}{\dots}& 0.85          &   0.84    &\multicolumn{2}{c}{$<$14.4}    & 6       \\
ISO\,32               & 39.2          &  4.0     &       67.5 & 22.7       &\multicolumn{2}{c}{$<$294.0}& T,T,T &\multicolumn{2}{c}{\dots}&\multicolumn{2}{c}{1.8}    &\multicolumn{2}{c}{$<$3}       & 7,8     \\
ISO\,102              & 92.1          &  1.6     &       98.6 &  6.8       &\multicolumn{2}{c}{$<$283.3}& T,T,T &\multicolumn{2}{c}{\dots}& 4.1           &   0.22    &\multicolumn{2}{c}{\dots}      & 9       \\
ISO\,160              & 48.2          &  3.5     &       71.1 &  5.1       &\multicolumn{2}{c}{$<$114.3}& T,T,T &\multicolumn{2}{c}{\dots}&\multicolumn{2}{c}{7.6}    &\multicolumn{2}{c}{\dots}      & 8       \\
\hline             
\multicolumn{15}{c}{\emph{Literature Sample}}          \\                                             
\hline             
L1521F-IRS            &        522    &   4      &\multicolumn{2}{c}{\dots}&  3712 &   52               &     1 & 1300     & 500          & \multicolumn{2}{c}{\dots} & 600    & 150\tablefootmark{a} & 10      \\
FW\,Tau\,A+B+C        &         30    &   4      &      33    &    4       &    70 &   40               &     1 &    5     & 1            & \multicolumn{2}{c}{\dots} &\multicolumn{2}{c}{$<$15}      & 11,12   \\
FN\,Tau               &       1755    &   4      &\multicolumn{2}{c}{\dots}&   816 &   16               &     1 &\multicolumn{2}{c}{\dots}&  36.5         &   5.0     &  31    &   1                  & 4,12    \\
CFHT\,12              &          2    &   1      &\multicolumn{2}{c}{\dots}&\multicolumn{2}{c}{$<$8}    &     1 &    4     & 1            & \multicolumn{2}{c}{\dots} &\multicolumn{2}{c}{$<$3}       & 6,4     \\
ZZ\,Tau\,IRS          &       2901    &   5      &\multicolumn{2}{c}{\dots}&  2922 &   26               &     1 &\multicolumn{2}{c}{\dots}& \multicolumn{2}{c}{\dots} & 106    &   2                  & 4       \\
J04381486+2611399     &         95    &   2      &\multicolumn{2}{c}{\dots}&    67 &   24               &     1 &\multicolumn{2}{c}{\dots}& \multicolumn{2}{c}{\dots} &   2.29 &   0.75               & 13      \\
CIDA\,7               &        330    &   2      &\multicolumn{2}{c}{\dots}&   342 &   19               &     1 &   38     & 8            & \multicolumn{2}{c}{\dots} &\multicolumn{2}{c}{$<$19}      & 4       \\
KPNO-10               &        160    &   2      &\multicolumn{2}{c}{\dots}&    82 &   26               &     1 &\multicolumn{2}{c}{\dots}& \multicolumn{2}{c}{\dots} &   8    &   1                  & 4       \\
IRAS\,04158+2805      &       1089    &   3      &\multicolumn{2}{c}{\dots}&  2953 &   25               &     1 &  407     & 41           & \multicolumn{2}{c}{\dots} & 110    &   5                  & 14,15   \\
J04202555+2700355     &        107    &   3      &\multicolumn{2}{c}{\dots}&   100 &   15               &     1 &\multicolumn{2}{c}{\dots}& \multicolumn{2}{c}{\dots} &   8    &   1                  & 4       \\
KPNO-3                &         23    &   4      &\multicolumn{2}{c}{\dots}&    33 &   12               &     1 &\multicolumn{2}{c}{\dots}& \multicolumn{2}{c}{\dots} &   6    &   1                  & 4       \\
MHO-6                 &        107    &   2      &\multicolumn{2}{c}{\dots}&   188 &    7               &     1 &\multicolumn{2}{c}{\dots}& \multicolumn{2}{c}{\dots} &  14    &   2                  & 4       \\
CFHT\,4               &        109    &   5      &\multicolumn{2}{c}{\dots}&\multicolumn{2}{c}{$<$150}  &     1 &   11     & 2            &   4.3         &   0.2     &   2.38 &   0.75               & 16,5,13 \\
ESO\_HA-559           &\multicolumn{2}{c}{$<$248}&     228.50 &   11.80    & 284.1 & 54.9               &     1 &\multicolumn{2}{c}{\dots}&  44.0         &   5.0     &\multicolumn{2}{c}{\dots}      & 17      \\
2M0444+2512           &        157.0  &  26.4    &\multicolumn{2}{c}{\dots}&\multicolumn{2}{c}{\dots}   &     2 &   10     & 1            &   9.0         &   0.2     &   5.20 &   0.30               & 5,6     \\ 
TWA 30B               &        65.7   &  1.8     &      55.6  &    2.1     &  48.4 & 3.0                &     3 &\multicolumn{2}{c}{\dots}& \multicolumn{2}{c}{\dots} &   0.83 &   0.07               & 18      \\ 
TWA 32                &        46.9   &  1.3     &      51.4  &    2.3     &  46.9 & 2.0                &     3 &\multicolumn{2}{c}{\dots}& \multicolumn{2}{c}{\dots} &   2.10 &   0.05               & 18      \\
TWA 34                &        24.5   &  1.2     &      18.7  &    1.8     &  17.2 & 2.5                &     3 &\multicolumn{2}{c}{\dots}& \multicolumn{2}{c}{\dots} &   0.54 &   0.06               & 18      \\
\hline
\end{tabular}
\tablefoot{
\tablefoottext{a}{Measured at $F_{1200}$ instead of $F_{1300}$.}
\tablebib{(T) This paper; (1) \citealt{bul14}; (2) \citealt{gui07}; (3) \citealt{liu15a}; (4) \citealt{and13}; (5) \citealt{ric14}; (6) \citealt{moh13}; (7) \citealt{pha11}; (8) \citealt{tes16}; (9) \citealt{ric12}; (10) \citealt{bou06}; (11) \citealt{and05}; (12) \citealt{bec90}; (13) \citealt{sch06}; (14) \citealt{and08}; (15) \citealt{mot01}; (16) \citealt{kle03}; (17) \citealt{bel11}; (18) \citealt{rod15a}}
}
\end{table*}

\subsubsection{Sample characteristics}
The final sample of 29 objects is a mixed group of VLMOs from different regions, dominated by objects in Taurus, but also including a few in $\rho$\,Ophiuchus (3), the TW Hydrae association (3), and Chamaeleon\,I (1). The sample ranges in mass from 0.03 to 0.2$\,M_{\odot}$. Most of the sources are likely to have ages between 1--3\,Myr, the exception being the 3 objects in TW Hydrae, which are probably significantly older. Our selection process, as defined above, essentially means that we pick objects with bright disks that have already been targeted by Herschel and by (sub)-mm campaigns, which is obviously not a well-defined category. Therefore, the sample is not complete down to a well-defined mass or luminosity limit. 

The same, however, can be said for all brown dwarf samples studied so far in the FIR and (sub)-mm domain. The sample used by \citet{har12} includes objects in nine nearby, diverse star-forming regions, with ages from 1 to 10\,Myr, handpicked for Herschel observations. The range in spectral types is slightly wider than ours, including objects that are earlier and later than our M4 and M8 limits. The objects analyzed by \citet{liu15} is based on the Harvey sample, plus some more objects observed by Herschel. Their sample is therefore, as the authors state, ``biased towards very low-mass substellar objects''. The studies by \citet{alv13} and \citet{bul14} cover almost all known disk-bearing brown dwarfs in $\rho$\,Ophiuchus and Taurus, respectively, but not all are detected. While these surveys are almost complete for the studied region, they suffer from the fact that the detection limit is not homogeneous. Such selection effects have to be taken into account when comparing results.

Three targets (CFHT-18, FW\,Tau, TWA\,32) have stellar or BD companions that fall within the photometric aperture of our far-IR measurements \citep{kon07,che90,shk11}. We keep them in the sample, but caution that binarity can have an effect on the SED and other measured parameters, depending on the brightness and color of the companion. {The existence of additional binary companions, including spectroscopic binaries, in the sample cannot be excluded, as no comprehensive binary survey was executed for the sample. Spectroscopic binaries among pre-main-sequence stars are, however, rare \citep[$\sim$6\% for low-mass stars in Taurus;][]{ngu12}.}

\subsection{Data reduction \& photometry}\label{sec:reduction}
\begin{figure*}[tbh]
\includegraphics[width=\textwidth]{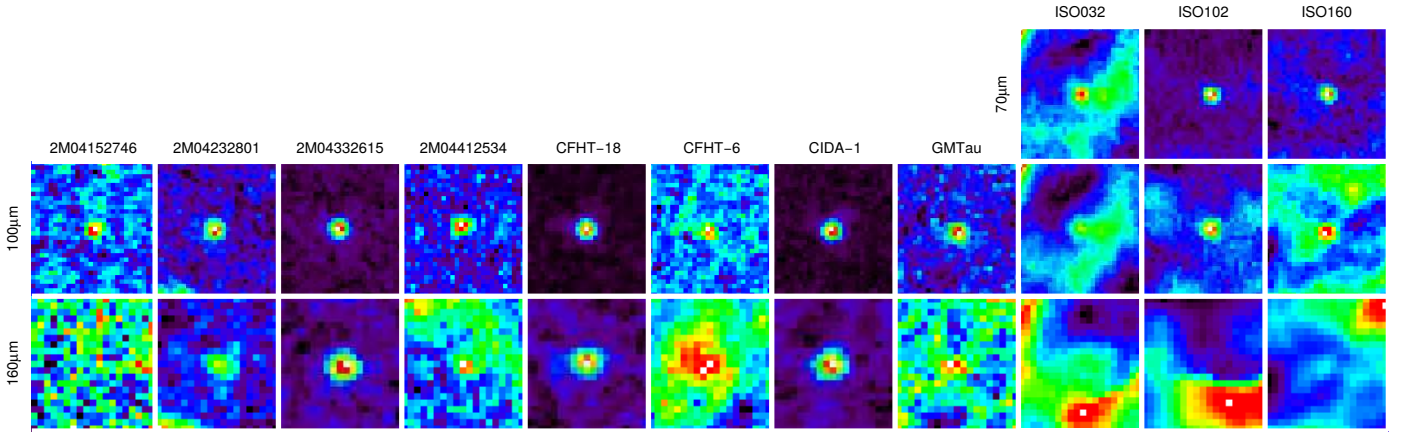}
\caption{\label{fig:stamps}PACS/Herschel observations centered on the individual targets. Postage stamps are 1\arcmin$\times$1\arcmin, North is up and east to the left.}
\end{figure*}

All Herschel photometry of our core sample of 11 targets rely on the \emph{phot project} \emph{level~3} automatic pipeline results (SPG 11.1.0), obtained using the \emph{Herschel Interactive Processing Environment} \citep[\emph{hipe};][]{ott10}. The pipeline combines all consecutive observations of the same target (usually 2) and performs high-pass filtering with a flux cut at 1.5 times the standard deviation of the flux per pixel and a filter width of 15, 15, and 32 readouts in the 70\,\mum\ (blue), 100\,\mum\ (green), and 160\,\mum\ (red) maps, respectively. The built-in MMT deglitching routine was applied and the output pixel scale was set to 2\arcsec, 3\arcsec, and 4\arcsec\ for the blue, green, and red filters, respectively. Smaller pixels scales have been tested for a subset of the reductions and did not lead to significantly different results. Image stamps of the fully reduced Herschel/PACS data are shown in Fig.~\ref{fig:stamps}.

Photometry was performed in \emph{IDL} using the routine {\tt aper.pro} to measure sky-subtracted flux values. If possible, our custom procedure recenters on the source and then performs aperture photometry. Aperture radii were set to the values used by \citet{alv13}, i.e., 5\farcs61, 6\farcs79, and 11\farcs39 in the blue, green, and red filters, respectively. Inner and outer sky radii were set to two and three times this value, respectively. If strong spatially extended emission close to the source was detected by eye (or indications of it in the curve of growth), aperture radii smaller by a factor of 1.5 (2M04152746) or 2 (2M04412534, CFHT-6, GM\,Tau, ISO32, ISO102, ISO160) were used to exclude as much as possible of the extended emission from leaking into the point-source photometry {and to ensure that the sky annuli provide a good estimate of the level of the background at the position of the target.}
Aperture corrections were applied according to the tabulated values in \citet{bal14}. Since the detected emission stems from cold dusty material in the circum-substellar disks, color correction factors of $-$3.4\%, 1.8\%, and 2.4\% in blue, green, and red bands were applied assuming an effective temperature of T=30\,K \citep{muller_pacs_2011}. The flux uncertainties introduced through the color correction factor by the -- a priory unknown -- disk temperatures are smaller than 2\%, 5\%, and 10\% in blue, green, and red for $T\ge30$\,K but can be much larger, particularly at 70\,\mum, for $T\lesssim20$\,K. Flux uncertainties were estimated from the scatter in 15 apertures of the same size as the science aperture, distributed on a circle around the source. Some apertures were excluded from the uncertainty estimate (2$\sigma$ outlier rejection) to lower the impact of spatial background variability at the position of the noise apertures.

As \citet{pop12} show, the high-pass filter applied by the automatic \emph{hipe/phot project} pipeline may compromise flux measurements close to extended emission. To measure the impact on our photometry, we compare our results with two alternative reductions: first, the \emph{hipe} pipeline as described above but without high-pass filtering and, second, the \emph{scanamorphos} pipeline that is designed to conserve large-scale emission \citep{rou13}.
We find agreement to within 1$\sigma$ between our results and those based on both alternative reductions for all targets. It appears, however, that most flux uncertainties in the \emph{scanamorphos} reduction are larger than their \emph{hipe}-reduced counterparts. This is likely because of gradients in the large-scale emission that cause artificially large uncertainties with the applied photometry routine. In the following, we use the fluxes derived from the automatic \emph{hipe}/\emph{phot project} pipeline.

\section{Results\label{sec:results}}
\subsection{Far-IR photometry \& SEDs\label{sec:phot}}
We detect all eleven targets at 100\,\mum. Seven of these were also detected at 160\,\mum, and we derive 3 sigma upper limits for the rest of the sample at 160\,\mum. Furthermore, all three attempted 70\,\mum\ observations led to detections. In Table \ref{tab:fluxes}, we list our results together with complementary literature values at 70\,\mum\ \citep{bul14}. 

Where available, we compare our far-IR measurements with previous observations of the same targets. \citet{bul14} measure $F_{160}$ fluxes for all eight Taurus targets of our core sample. Most of our fluxes in Table~\ref{tab:fluxes} agree to within 1$\sigma$ with the \citeauthor{bul14} values. Two targets (CIDA-1 and 2M04232801) show deviations of 1.5$\sigma$ and 2.2$\sigma$, respectively. The discrepancy can be explained by source variability, which is an intrinsic feature of young low-mass stars even at far-IR wavelengths \citep{bil12}. Despite the fact that we compare optical, far-IR, and sub-mm photometry from different epochs, the observed variability of up to $>$20\% at 70--160\,\mum\ does not severely affect our analysis because far-IR fluxes enter our analysis mostly logarithmically (e.g., Eq.~\ref{eq:slopes}), i.e., as an uncertainty of $\sim0.08$\,dex.
 We find agreement for ISO\,32 and ISO\,160 to within the uncertainties with the values derived by \citet{alv13} in all three PACS bands. For ISO\,102, \citet{alv13} measure $F_{70}=80.1\pm5.6$ and $F_{100}=48.4\pm19.1$, which are both significantly smaller than our measurements (2$\sigma$ and 2.5$\sigma$, respectively). Part of the difference can be explained by the different effective temperature assumed for the color correction. The difference between a color correction at $T_\mathrm{eff}=30$\,K and the temperature of $T_\mathrm{eff}=1000$\,K assumed by \citet{alv13} is $\sim$2\%, $\sim$5\%, and $\sim$10\% in the blue, green, and red bands. As Herschel fluxes are dominated by emission from the circumstellar dust at low temperatures, we assume that 30\,K is a better representation of the color correction. The remaining discrepancy may be due to systematic biases of the different photometry strategies, which are likely enhanced by the strong nebular background in the surroundings of ISO\,102 or by source variability. As the signal-to-noise ratio of our pointed observations is higher than that of the scan maps produced by \citet{alv13}, we use our photometry in the following.

Using literature data for 18 additional targets, we compiled the largest and mostly complete collection of very low-mass stars and brown dwarfs that were detected at both far-IR and (sub)-mm wavelengths (Tab.~\ref{tab:fluxes}). We present a sample of 29 targets that were detected in at least one of the Herschel PACS bands (28 detections at 70\,\mum, and 22 at 160\,\mum). Twenty-eight of these have at least one detection at (sub)-mm wavelengths between 850\,\mum\ and 1300\,\mum. For one target, CFHT-18, only an upper limit at 1300\,\mum\ was found in the literature.

\begin{figure}[tb]
\centering
\includegraphics[width=\columnwidth]{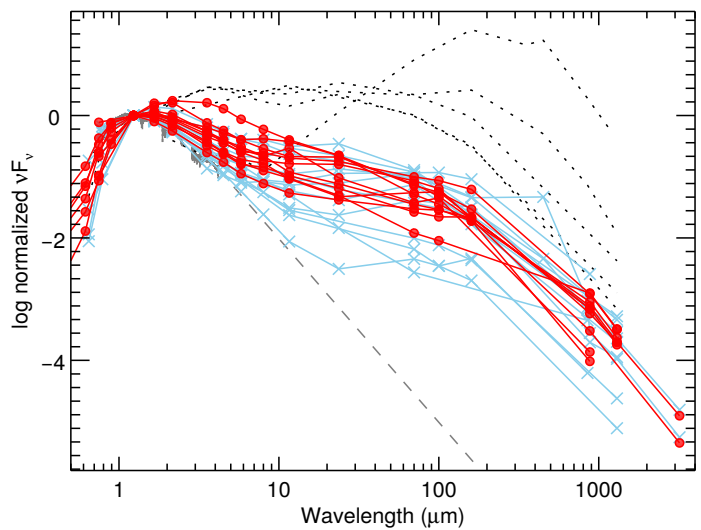}
\caption{\label{fig:normalized_SEDs} 
 Dereddened SEDs normalized at $J$ band. Red curves show the SEDs of the original sample, and blue lines show the literature SEDs. Targets that were excluded from further discussion (see Sect.~\ref{sec:phot}) are shown with dotted lines. For comparison, the expected photospheric emission of an M5 star is shown in dashed gray \citep{cus05}.}
\end{figure}

We compile spectral energy distributions (SED) between optical and millimeter bands from the literature (Appendix \ref{sec:appA}). For easy comparison, all SEDs scaled to their $J$-band flux are shown in Fig.~\ref{fig:normalized_SEDs}.

The SEDs of five targets (IRAS\,04158+2805, L1521F-IRS, ZZ\,Tau\,IRS, 2MASS\,J04381486+2611399, and TWA\,30B, i.e., all targets with a horizontal or positive slope between 10 and 100\,\mum\ in Fig.~\ref{fig:normalized_SEDs}), show indications of a dense circumstellar envelope such as that of Class\,I sources or an edge-on disk. Owing to the different nature and uncertain photometry of these targets, they are excluded from subsequent disk parameter estimates. 

\begin{figure}[tb]
\centering
\includegraphics[width=\columnwidth]{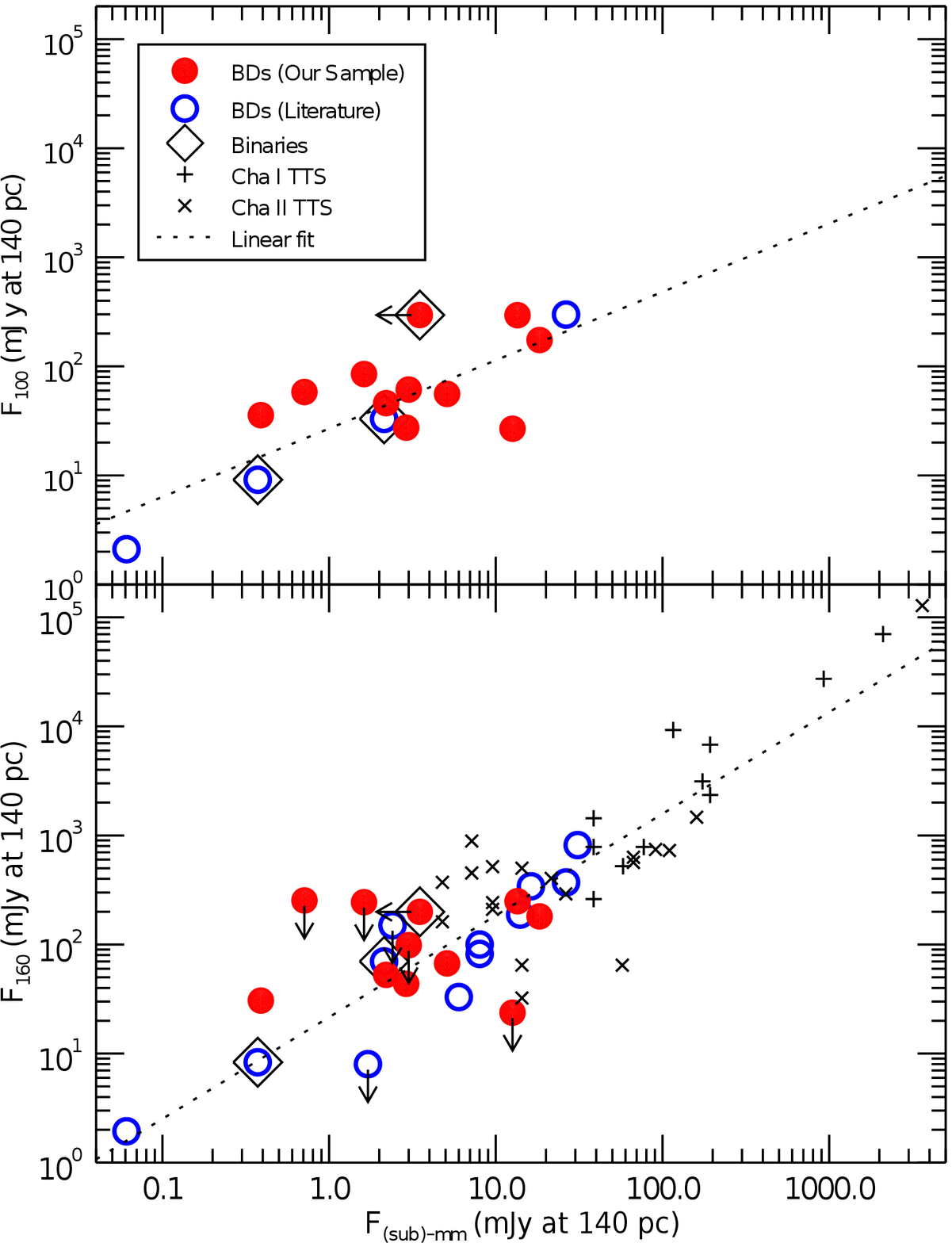}
\caption{\label{fig:fFIR_vs_fmm}
  $F_{100}$ and $F_{160}$ as a function of mm fluxes. Millimeter fluxes are identical to $F_{1300}$ or were converted from other wavelengths according to eq.~(\ref{eq:mmconversion}). To compare, T Tauri stars in Chamaeleon\,I and II are shown as plus and cross symbols in the bottom panel \citep{win12,spe13}. The dotted lines show linear fits to all points except upper limits and binaries.}
\end{figure}

For our VLMO sample, we notice a strong correlation between FIR and (sub)-mm fluxes. Fig.~\ref{fig:fFIR_vs_fmm} shows the correlations of $F_{100}$ and $F_{160}$ with $F_\mathrm{mm}$ (similar for $F_{70}$).
The latter is equal to $F_{1.3\,\mathrm{mm}}$, where available, or has been converted from $F_{850}$ or $F_{890}$ (see Table~\ref{tab:fluxes}) according to
\begin{equation}\label{eq:mmconversion}
  F_\mathrm{mm}=\left(\frac{D}{140\,\mathrm{pc}}\right)^2\left(\frac{\nu}{230\,\mathrm{GHz}}\right)^{-3}\cdot F_{\nu}
.\end{equation}
This equation has been derived assuming that the (sub)-mm flux is dominated by optically thin emission and can be approximated with the Rayleigh-Jeans law ($F_{\nu}\propto\nu^2$). We further use a linear opacity dependence ($F_\nu\propto\kappa_\nu$ with $\kappa_\nu \propto \nu^{\beta}$ for $\beta=1$, see also Sect.~\ref{sec:diskmasses}) and scale all measurements to a distance of $D=140$\,pc.

A Spearman's rank test returns correlation parameters of 0.61, 0.62, and 0.94 for correlations of $F_\mathrm{mm}$ with $F_{70}$, $F_{100}$, and $F_{160}$, respectively, taking into account detections in the whole VLM sample and excluding binary stars. The respective significances are 6$\times$10$^{-3}$, 3$\times$10$^{-2}$, and 8$\times$10$^{-7}$, i.e., the measured correlations are significantly different from zero. The measured slopes of best linear fits to the data (in log-log space, considering the same selection as for the correlation coefficients) are $m_{70}\!=\!1.33\pm0.76$, $m_{100}\!=\!1.43\pm0.63$, and $m_{160}\!=\!1.33\pm0.93$. The observed correlations also seem to hold for more massive T Tauri stars, which are shown with black symbols in Fig.~\ref{fig:fFIR_vs_fmm}~ \citep{win12,spe13}. Their mm fluxes have been derived from $F_{500}$ using Eq.~(\ref{eq:mmconversion}) as well. We discuss the physical reason for the observed correlation in Sect.~\ref{sec:fFIR_vs_fmm}.

\subsection{Spectral slopes}\label{sec:slopes}
As a diagnostic for far-infrared fluxes of our young brown dwarf sample, we derive spectral slope indices following \citep{ada87}:
\begin{equation}\label{eq:slopes}
 \alpha_{\lambda_1 - \lambda_2} = \frac{\log(\lambda_1 F_{\lambda_1}) - \log(\lambda_2 F_{\lambda_2})}{\log(\lambda_1) - \log(\lambda_2)}\quad.
\end{equation}
Such indices provide model-free estimates of the strength of far-IR dust emission, independent of stellar luminosity. In Table~\ref{tab:diskmasses}, we present the slopes between the $J$ band ($\lambda_0$=1.24\,\mum), which is dominated by the stellar photosphere, and the Herschel/PACS bands $\alpha_{J-70}$, $\alpha_{J-100}$ and $\alpha_{J-160}$.
\begin{table*}
\caption{Far-IR slopes and disk masses\label{tab:diskmasses}}
\centering
\small
\begin{tabular}{lccccc}
\hline\hline
{} &
{} &
{} &
{} &
$\log(M_\mathrm{disk}/M_\odot)$ &
$\log(M_\mathrm{disk}/M_\odot)$ \\
{Name} &
{$\alpha_{J-70}$} &
{$\alpha_{J-100}$} &
{$\alpha_{J-160}$} &
($T=25(L/L_\odot)^{1/4}$\,K) &
($T=20$\,K)\\
\hline
\multicolumn{6}{c}{\emph{Our Sample}}\\                                       
\hline
  2M04152746              & $-$1.10     &  $-$1.07& $<$$-$1.09 & $-$2.39\,$\pm$\,0.10              & $-$2.63       \,$\pm$\,  0.10   \\
  2M04232801              & $-$0.87     &  $-$0.81&    $-$0.79 & $-$2.70\,$\pm$\,0.13              & $-$3.03       \,$\pm$\,  0.13   \\
  2M04332615              & $-$0.83     &  $-$0.81&    $-$0.82 & $-$2.34\,$\pm$\,0.10              & $-$2.47       \,$\pm$\,  0.10   \\
  2M04412534              & $-$0.57     &  $-$0.56&    $-$0.57 & $-$2.81\,$\pm$\,0.12              & $-$3.39       \,$\pm$\,  0.12   \\
  CFHT6                   & $-$0.90     &  $-$0.87&    $-$0.78 & $-$2.86\,$\pm$\,0.15              & $-$3.27       \,$\pm$\,  0.15   \\
  CFHT18                  & $-$0.64     &  $-$0.69&    $-$0.80 & $<$\,$-$3.06                      & $<$\,$-$3.19                    \\
  CIDA1                   & $-$0.65     &  $-$0.66&    $-$0.72 & $-$2.45\,$\pm$\,0.13              & $-$2.60       \,$\pm$\,  0.13   \\
  GMTau                   & $-$0.74     &  $-$0.76&    $-$0.81 & $-$3.71\,$\pm$\,0.30              & $-$4.24       \,$\pm$\,  0.30   \\
  ISO032                  & $-$0.81     &  $-$0.71& $<$$-$0.43 & $-$3.49\,$\pm$\,0.10              & $-$3.98       \,$\pm$\,  0.10   \\
  ISO102                  & $-$0.61     &  $-$0.63& $<$$-$0.44 & $-$3.16\,$\pm$\,0.09              & $-$3.62       \,$\pm$\,  0.09   \\
  ISO160                  & $-$0.72     &  $-$0.65& $<$$-$0.59 & $-$2.84\,$\pm$\,0.09              & $-$3.35       \,$\pm$\,  0.09   \\
\hline                
\multicolumn{6}{c}{\emph{Literature Sample}}\\    
\hline                
  L1521F-IRS              & $>$1.26     & \dots   &    $>$1.28 & (\dots)\tablefootmark{a}          & (\dots)\tablefootmark{a}        \\
  FW\,Tau\,A+B+C          & $-$1.34     & $-$1.29 &    $-$1.11 & $-$3.43\,$\pm$\,0.13              & $-$3.52       \,$\pm$\,  0.13   \\
  FN\,Tau                 & $-$0.53     & \dots   &    $-$0.76 & $-$2.29\,$\pm$\,0.16              & $-$2.24       \,$\pm$\,  0.16   \\
  CFHT\,12                & $-$1.46     & \dots   & $<$$-$1.10 & $-$3.09\,$\pm$\,0.14              & $-$3.62       \,$\pm$\,  0.14   \\
  ZZ\,TauIRS              &    0.19     & \dots   &    $-$0.01 & (\dots)\tablefootmark{a}          & (\dots)\tablefootmark{a}        \\
  J04381486+2611399       & $-$0.01     & \dots   &    $-$0.25 & (\dots)\tablefootmark{a}          & (\dots)\tablefootmark{a}        \\
  CIDA\,7                 & $-$0.57     & \dots   &    $-$0.64 & $-$2.43\,$\pm$\,0.13              & $-$2.64       \,$\pm$\,  0.13   \\
  KPNO-10                 & $-$0.64     & \dots   &    $-$0.84 & $-$2.59\,$\pm$\,0.10              & $-$2.83       \,$\pm$\,  0.10   \\
  IRAS\,04158+2805        &    0.19     & \dots   &       0.20 & (\dots)\tablefootmark{a}          & (\dots)\tablefootmark{a}        \\
  J04202555+2700355       & $-$0.54     & \dots   &    $-$0.63 & $-$2.44\,$\pm$\,0.10              & $-$2.83       \,$\pm$\,  0.10   \\
  KPNO\,3                 & $-$0.81     & \dots   &    $-$0.77 & $-$2.49\,$\pm$\,0.12              & $-$2.96       \,$\pm$\,  0.12   \\
  MHO-6                   & $-$0.79     & \dots   &    $-$0.71 & $-$2.38\,$\pm$\,0.11              & $-$2.59       \,$\pm$\,  0.11   \\
  CFHT\,4                 & $-$0.60     & \dots   & $<$$-$0.60 & $-$3.00\,$\pm$\,0.18              & $-$3.35       \,$\pm$\,  0.18   \\
  ESO\,HA-559             & $<$$-$0.42  & $-$0.49 &    $-$0.49 & $-$2.08\,$\pm$\,0.10              & $-$2.41       \,$\pm$\,  0.10   \\
  2M0444+2512             & $-$0.51     & \dots   &      \dots & $-$2.67\,$\pm$\,0.09              & $-$3.02       \,$\pm$\,  0.09   \\
  TWA30B                  &    0.00     & $-$0.12 &    $-$0.23 & (\dots)\tablefootmark{a}          & (\dots)\tablefootmark{a}        \\
  TWA32                   & $-$1.14     & $-$1.11 &    $-$1.12 & $-$3.75\,$\pm$\,0.09              & $-$4.16       \,$\pm$\,  0.09   \\
  TWA34                   & $-$1.25     & $-$1.29 &    $-$1.28 & $-$4.42\,$\pm$\,0.10              & $-$4.95       \,$\pm$\,  0.10   \\
  \hline
\end{tabular}
\tablefoot{
  \tablefoottext{a}{These targets were excluded from further evaluation of their disk properties for reasons explained in Sect.~\ref{sec:phot}.}
  }
\end{table*}

\begin{figure*}[tb]
\centering
\includegraphics[width=1\textwidth]{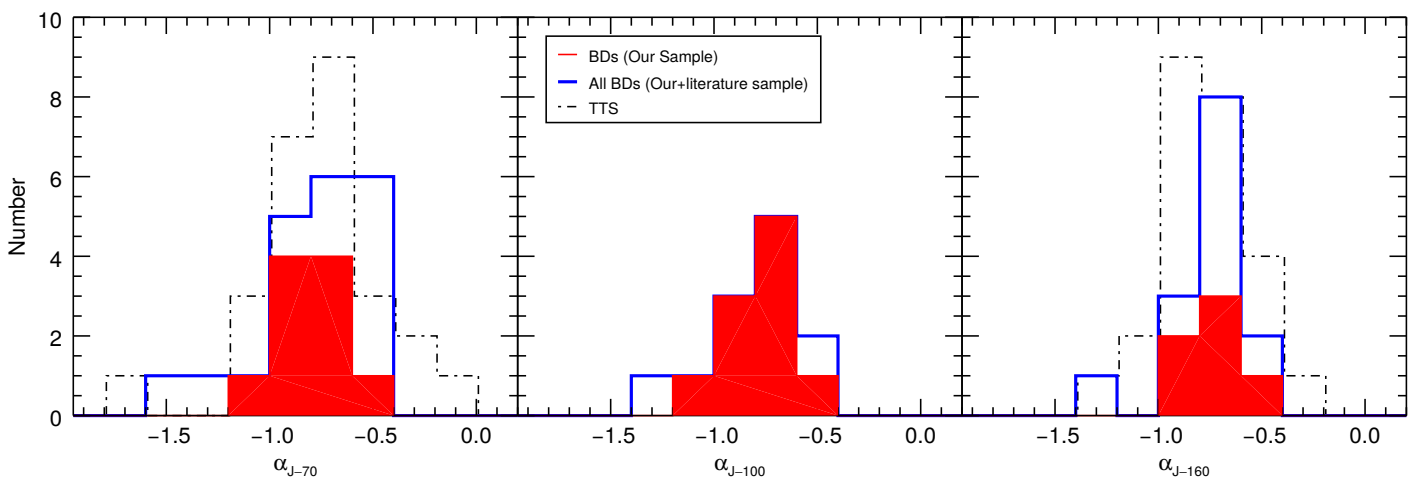}
\caption{\label{fig:IRslope_hist} Histograms of infrared slopes $\alpha_{J-70}$, $\alpha_{J-100}$, and $\alpha_{J-160}$. At $\alpha_{J-70}$ and $\alpha_{J-160}$, our brown dwarf samples are compared to T Tauri stars in Chamaeleon\,I and II \citep{win12,spe13}.} 
\end{figure*}

A histogram of the slope distribution is shown in Fig.~\ref{fig:IRslope_hist}.
We find that the NIR-FIR slopes of the VLMOs in our samples range mostly between $-$1.0 and $-$0.5. The median is $-0.72$,$-0.71$, and $-0.76$ for $\alpha_{J-70}$, $\alpha_{J-100}$, and $\alpha_{J-160}$, respectively, with a 1-sigma spread of $\sim$0.2. For comparison, we calculated the same slopes for a literature sample of T Tauri stars from \citet{win12} and \citet{spe13}, also shown in Fig.~\ref{fig:IRslope_hist} as a dash-dotted histogram. The range found for T Tauri stars is very similar to the values quoted for our BD sample, while the medians are marginally smaller ($-0.77$,$-0.79$ for $\alpha_{J-70}$,$\alpha_{J-160}$), but still well within the standard deviation. Thus, BDs and T Tauri stars have similar spectral slopes in the FIR.

We also calculate the same slopes for the Class\,II Taurus sample by \citet{bul14}, which partially overlaps with ours but covers a slightly larger spectral type range. Again the range of the slopes and the peak of the distribution are similar to our results (median $-0.71$,$-0.73$ for $\alpha_{J-70}$,$\alpha_{J-160}$). Since the Bulger et al.  sample covers a large percentage of all known low-mass stars in Taurus, the similarity provides reassurance that the spectral slope distribution of our BD sample is not significantly affected by selection biases.

\begin{figure}[tb]
\centering
\includegraphics[width=1\columnwidth]{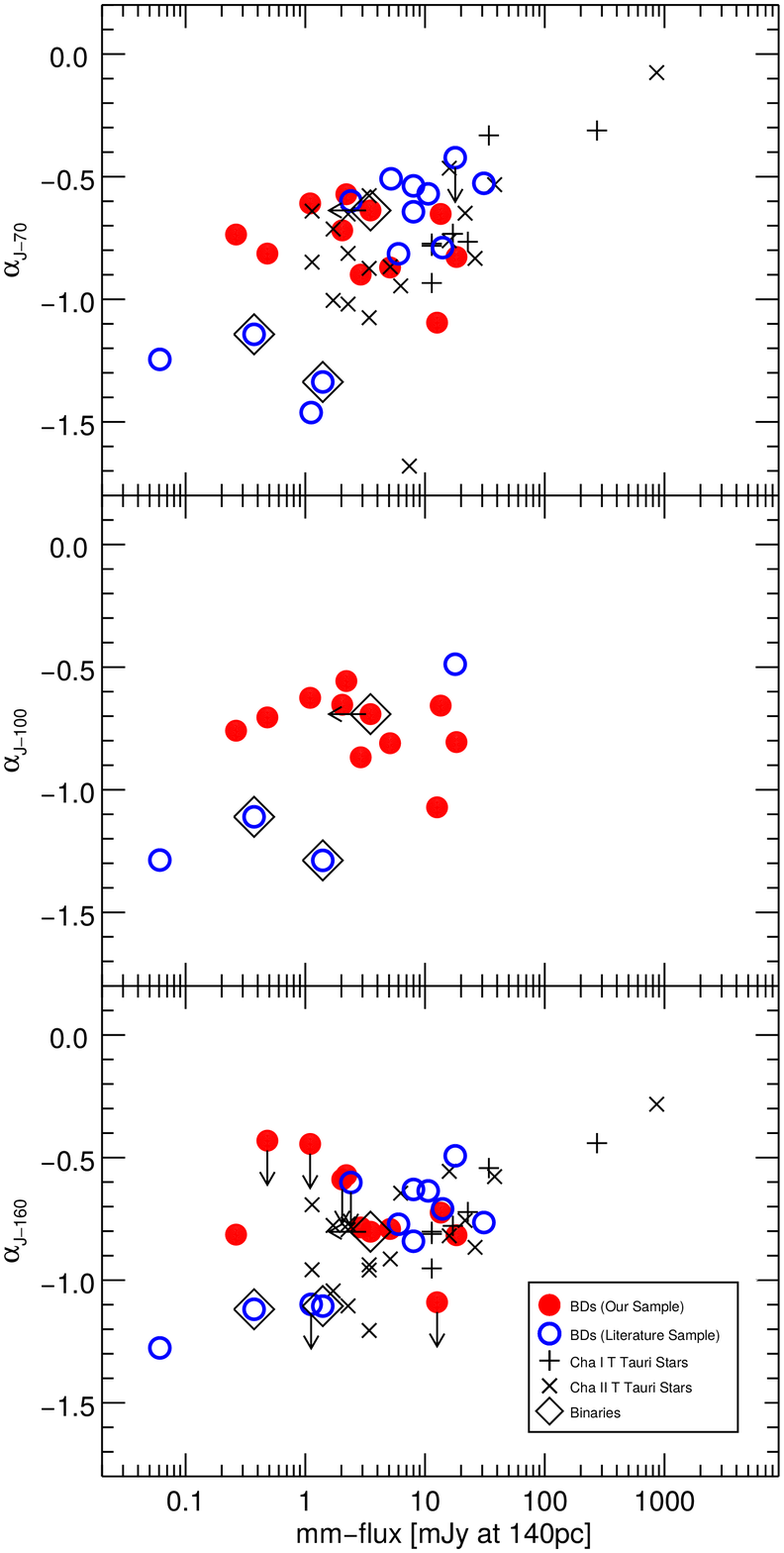}
\caption{\label{fig:IRslope_vs_mm} Infrared slope as a function of mm flux. Millimeter fluxes are identical to $F_{1300}$ or were converted from other wavelengths according to eq.~(\ref{eq:mmconversion}). Binaries are indicated with diamonds. For comparison, data from more massive T\,Tauri stars in the Chamaeleon star-forming region are shown (Cha\,I, \citealt{win12}; Cha\,II, \citealt{spe13}).} 
\end{figure}

In Fig.~\ref{fig:IRslope_vs_mm}, we compare $\alpha_{J-70}$, $\alpha_{J-100}$, and $\alpha_{J-160}$ with the mm-flux $F_\mathrm{mm}$ (eq.~\ref{eq:mmconversion}), which is expected to be proportional to the disk mass (see Sect.~\ref{sec:diskmasses}). 
For comparison with higher mass counterparts, we again use the slopes derived from the T Tauri surveys by \citeauthor{win12} and \citeauthor{spe13} and mm fluxes converted from $F_{500}$.
The numbers are added to Fig.~\ref{fig:IRslope_vs_mm} as black crosses. The bulk of the data points for VLMOs and T Tauri stars occupy the same region in these diagrams, without obvious trends, and a Spearman's rank test cannot rule out that BD slopes are uncorrelated with mm fluxes (significances of 0.27, 0.65, and 0.25 for $\alpha_{J-70}$, $\alpha_{J-100}$, and $\alpha_{J-160}$, respectively, based on detections in the whole VLM sample exluding binary stars).

\subsection{Disk masses}\label{sec:diskmasses}
We convert (sub)-mm fluxes to disk masses using
\begin{equation}\label{eq:eq2}
M_\mathrm{d} = \frac{D^2F_\nu}{\kappa_\nu B_\nu(T_\mathrm{d})}\quad,
\end{equation}
with distance $D$ from Table~\ref{tab:stellar}, $\kappa_\nu = \kappa_f\left(\frac{\nu}{\nu_f}\right)^\beta$ for $\beta=1$ and $\kappa_{f=230\mathrm{GHz}}=0.02$\,cm$^2$g$^{-1}$, assuming a gas:dust ratio of 100:1. We note that choosing $\beta=2$ instead of $\beta=1$ decreases disk mass estimates of targets observed at $F_{890}$ by $\sim$0.2\,dex. When available, $F_\nu=F_{1300}$, otherwise $F_{890}$ or $F_{850}$ are used (see Table~\ref{tab:fluxes}).
The dust temperature $T_\mathrm{d}$ is calculated following the prescription by \citet{and13}
 \begin{equation}\label{eq:eq2a}
 \langle T_\mathrm{d}\rangle = 25\,(L_\star/L_\odot)^{1/4}\,\mathrm{K}
,\end{equation}
which yields dust temperatures between $\sim$7 and 20\,K. This scaling of the dust temperature with the stellar luminosity appears to be more realistic than using fixed values between 10 and 20\,K as in previous brown dwarf disk surveys (e.g., $T=15$\,K in \citealt{sch06}, 20\,K in \citealt{bro14}); see Sect.~\ref{sec:Md_vs_fmm} for further justification. In Table~\ref{tab:diskmasses}, we list the derived disk masses according to eqs.~(\ref{eq:eq2},\ref{eq:eq2a}). For reference, we also include disk masses with $T_\mathrm{d}=20$\,K.
 
\begin{figure}[tb]
\centering
\includegraphics[width=\columnwidth]{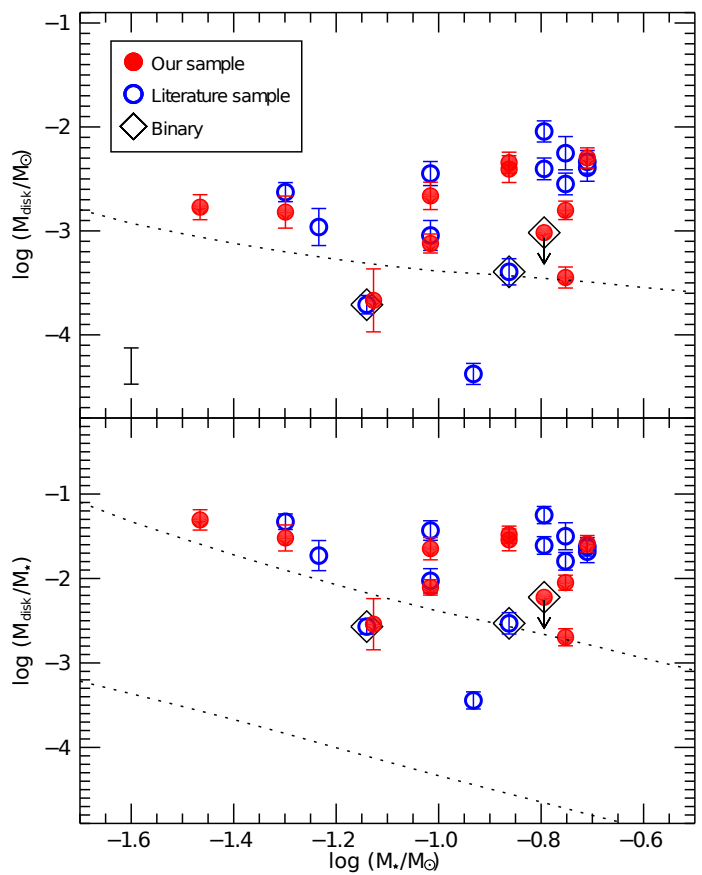}
\caption{\label{fig:disk_mass}Disk mass, derived from (sub)-mm fluxes, in units of solar masses (top panel) and stellar mass (bottom panel) as a function of the mass of the parent body for all targets in this paper. Binaries are highlighted with diamonds. A conservative estimate of disk mass uncertainty arising from unknown individual distances is shown in the bottom left of the top panel, assuming a conservative $\Delta$dist=$\pm$10\% \citep[cf.][who estimate a ``depth'' of Taurus of 20\,pc]{tor07}. Dotted lines show the typical sensitivity limits derived from eq.~\ref{eq:eq2} for targets in Taurus ($F_\mathrm{1300,lim}$$\approx$1\,mJy, dist=140\,pc) and TW\,Hya ($F_\mathrm{1300,lim}$$\approx$0.05\,mJy, dist=50\,pc), using BCAH98 isochrones to convert mass to luminosity.}
\end{figure}

Fig.~\ref{fig:disk_mass} shows the distribution of disk masses inferred from (sub)-mm fluxes as a function of stellar host mass.
With disk masses at around 1\% of the stellar mass, we see that the current sample shows disk masses mostly at the upper edge of the disk mass distribution in the respective stellar mass range. This is a selection effect because the required detection at (sub)-mm wavelengths prefers targets with the largest reservoirs of circumstellar dust.
The figure confirms that relative disk masses for brown dwarfs and stars are in the same range, i.e., the disk mass scales with object mass.

\section {Disk models}\label{sec:models}
To investigate the implications of our observational results on the physical properties of VLMO disks, such as flaring and mass, we ran a small grid of disk models. In our models, each disk is characterized by its dust mass, inner and outer radii, and the density distribution of dust in the radial ($R$) and vertical ($z$) directions. If $\Sigma$ is the column density and \Hp\ the pressure scale height at a given $R$, then the density is given by
\begin{equation}
 \rho (R,z)= \rho(R,z=0) \exp{(-z^2/2H_p^2)}
,\end{equation}
where
\begin{equation}
 \rho(R,z=0)= \frac{\Sigma(R)}{\sqrt{2\pi}\,H_p(R)}
,\end{equation}
where $\Sigma \propto R^{-p}$ and a normalization are defined by the disk mass. The dust pressure scale height \Hp\ has a well-defined physical meaning when the disk is in hydrostatic equilibrium and dust and gas are well mixed. This computation involves an iterative solution of the radiation transfer, as \Hp\ depends on the temperature structure, which, in turn depends on the disk geometry (and thus \Hp). Moreover, the dust is likely not well mixed, if significant differential settling occurs. Therefore, in most of the recent papers \citep[e.g.,][]{sic15,liu15}, \Hp\ is assumed to be a power-law function of $R$,
\begin{equation}
 H_p(R)= H_p(R_0) \>\> (R/R_0)^\xi
,\end{equation}
where $H_p(R_0)$ is the pressure scale height at the fiducial radius $R_0$ and $\xi$ is a free parameter, which is often defined in the literature as the flaring index. We note that in fully flared, hydrostatic equilibrium disks $\xi \sim 1.3$ and in flat disks $\xi\sim 1.1$. The shape of the disk is defined by the combination of $H_p(R_0)$ and $\xi$, which determine the emission at all wavelengths once all other parameters are fixed \citep{bul14,sic15}. In summary, a disk model is defined by six parameters: the disk mass \Mdisk, inner and outer radii \Ri\ and \Rout, the slope of the surface density profile $p$, and the two parameters $H_p(R_0)$, $\xi$.

In addition, one needs to fix the stellar parameters (mass, luminosity, and radius), and the dust opacity. We adopt a dust model, which is a mixture of 7\% (in volume) silicates \citep[{amorphous Mg-rich grains with olivine stoichiometry}, optical constant from][]{jae94}, 21\% carbonaceous materials \citep[aC\_ACH2, optical constants from][]{zub96}, 42\% water ice \citep{war84}, and 30\% vacuum. For most models, the grain size ranges from a minimum $a_\mathrm{min}=0.2$\,\mum\ to $a_\mathrm{max}=150$\,\mum\ with slope 3.5. The corresponding opacity is 2\,\cmg\ at 1300\,\mum, 150\,\cmg\ at 160\,\mum,\ and 700\,\cmg\ at 1\mum.
The opacity at 1.3 mm is the same as the value we used to calculate disk masses in Sect.~\ref{sec:diskmasses}. The opacities at 1.3\,mm and 890\,\mum\ are also simliar to those used by \citet{and13} and many other studies \citep[e.g.,][]{ric14}.

We compute the temperature structure and the emission at selected wavelengths in the FIR and (sub)-mm using the radiation transfer code RADMC-3D\footnote{\url{http://www.ita.uni-heidelberg.de/~dullemond/software/radmc-3d/}} (version 0.39; C.P. Dullemond). We consider two different central objects; one object, BD in the following, with $L_\star=0.03 L_\odot$, $M_\star=0.05 M_\odot$, and $T_\star=2800$\,K, and a more luminous object, TTS in the following, with $L_\star=0.15 L_\odot$, $M_\star=0.2 M_\odot$,  and $T_\star=3150$\,K. All our targets and most of those from the literature have luminosities and effective temperatures roughly within this range. We include in the calculations the heating due to the interstellar radiation field (IRF) as in \citet{dra11}. As the FIR and (sub)-mm fluxes depend only weakly on the disk inclination as long as $i$ is reasonably small ($i$$\lesssim$60--65\,deg), we fix $i=0$ in all cases.

Table~\ref{tab:tab4} gives the values of the disk parameters for each model.
\begin{table}
\caption{Model parameters\label{tab:tab4}}
\centering
\tiny
\begin{tabular}{lcccc|c}
\hline\hline
  {}  &
  {log($M_\mathrm{disk}$)} &
  {$H_p\,(10$AU)} &
  {$\xi$}  &
  {R$_\mathrm{out}$} &
  {$H_p\,(100$AU)}\tablefootmark{b}\\
  {Model\tablefootmark{a}} &
  {[M$_\odot $]} &
  {[AU]}&
  {}&
  {[AU]}&
  {[AU]}\\
\hline
\multicolumn{6}{c}{\emph{BD}}\\
\hline
mod1a       &   $-$4.5 &   1.5 & 1.35 &  50 & 33.6 \\ 
mod1b       &   $-$4.5 &  0.15 & 1.25 &  50 &  2.7 \\ 
\smallskip
mod1c       &   $-$4.5 &  0.15 & 1.10 &  50 &  2.0 \\ 
mod2a       &   $-$3.5 &  1.5  & 1.35 &  50 & 33.6 \\ 
mod2b       &   $-$3.5 &  0.15 & 1.25 &  50 &  2.7 \\ 
mod2b$^1$   &   $-$3.5 &  0.15 & 1.25 &  50 &  2.7 \\ 
mod2b$^2$   &   $-$3.5 &  0.15 & 1.25 &  50 &  2.7 \\ 
mod2b,small &   $-$3.5 &  0.15 & 1.25 &  30 &  2.7 \\ 
mod2b,large &   $-$3.5 &  0.15 & 1.25 & 100 &  2.7 \\ 
\smallskip
mod2c       &   $-$3.5 &  0.15 & 1.10 &  50 &  2.0 \\ 
mod3a       &   $-$2.5 &  1.5  & 1.35 &  50 & 33.6 \\ 
mod3b       &   $-$2.5 &  0.15 & 1.25 &  50 &  2.7 \\ 
mod3c       &   $-$2.5 &  0.15 & 1.10 &  50 &  2.0 \\ 
\hline
\multicolumn{6}{c}{\emph{TTS}}\\
\hline
mod2a       &   $-$3.5 &  1.5  & 1.35 & 100 & 33.6 \\ 
mod2b       &   $-$3.5 &  0.15 & 1.25 & 100 &  2.7 \\ 
\smallskip
mod2c       &   $-$3.5 &  0.15 & 1.10 & 100 &  2.0 \\ 
mod3a       &   $-$2.5 &  1.5  & 1.35 & 100 & 33.6 \\ 
mod3b       &   $-$2.5 &  0.15 & 1.25 & 100 &  2.7 \\ 
\smallskip
mod3c       &   $-$2.5 &  0.15 & 1.10 & 100 &  2.0 \\ 
mod4a       &   $-$1.8 &  1.5  & 1.35 & 100 & 33.6 \\ 
mod4b       &   $-$1.8 &  0.15 & 1.25 & 100 &  2.7 \\ 
mod4c       &   $-$1.8 &  0.15 & 1.10 & 100 &  2.0 \\ 
\hline
\end{tabular}
\tablefoot{\tablefoottext{a}{mod2b$^1$ and mod2b$^2$  are computed with different grain size distribution: in the first case $a_\mathrm{max}$=2.4\,cm ($\kappa_\nu=2.5$\,\cmg\ at 1300\,\mum, 14\,\cmg\ at 160\,\mum,\ and 56\,\cmg\ at 1\,\mum), in the second $a_\mathrm{max}$=1.5\,\mum\ ($\kappa_\nu=1.5$\,\cmg\ at 1300\,\mum, 94\,\cmg\ at 160\,\mum,\ and 4.5$\times$$10^3$\,\cmg\ at 1\,\mum).}
  \tablefoottext{b}{Scale heights of the disk profile at $R=100$\,AU as a reference for comparison with previous studies.}
}
\end{table}
In all cases, $p=1$ and $R_\mathrm{i}=0.1$\,AU. The values of $H_p(R_0)$ and $\xi$ were chosen to encompass a large range of possibilities, from very flat to very flared disks. We note that fully flared disks, when computed using a two-layer disk model as in \citet{chi97,dul01} have $H_p(10\mathrm{\,AU})\sim$1.5\,AU and $\sim$1\,AU for the BD and TTS, respectively, with $\xi\,\sim\,1.34$--1.30. Models with higher values of $H_0$ and/or $p$ are difficult to justify. In Table~\ref{tab:tab4} we give total disk masses, with the usual assumption of a gas-to-dust mass ratio of 100, even if the disk models only depend on the dust mass. For an easier comparison with what is often used by other authors, for each model we give the value of \Hp\ at $R=100$\,AU  in the last column.

\begin{figure}[tb]
 \centering
  \includegraphics[width=\columnwidth]{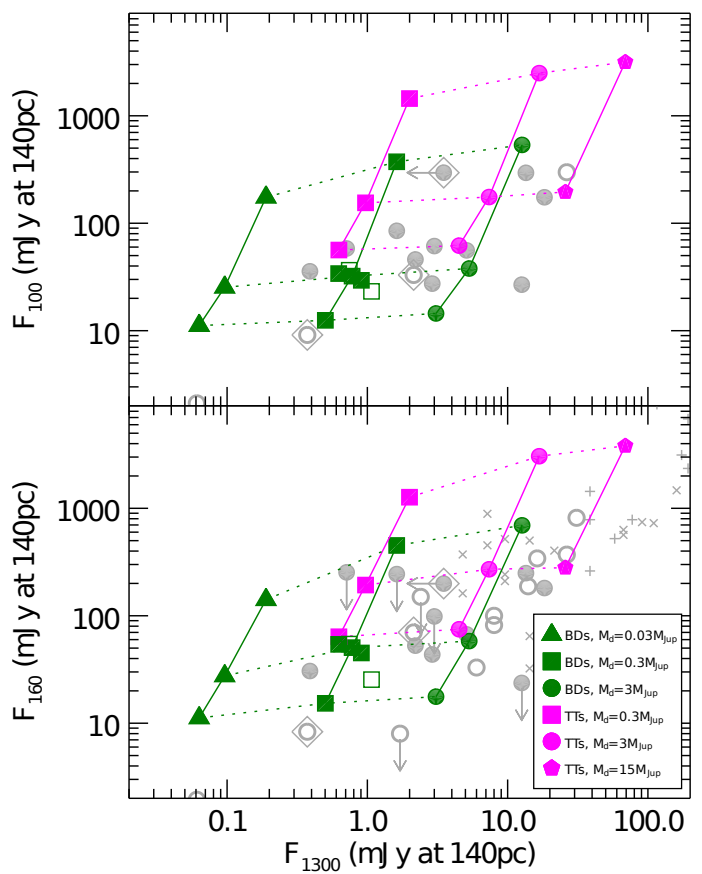}
 \caption{{Model-predicted 100\,\mum\ (top panel) and 160\,\mum\ fluxes (bottom) vs. 1300\,\mum\ flux.} Each colored point represents a model. Green symbols indicate the BD models, magenta symbols the TTS models. Different symbols refer to different disk masses: triangles for \Mdisk=3$\times$$10^{-5}$\Msun, squares for \Mdisk=3$\times$$10^{-4}$\Msun, dots for \Mdisk=3$\times$$10^{-3}$\Msun, pentagons for \Mdisk=1.5$\times$$10^{-2}$\Msun. Models for the same \Mdisk\ and different flares are connected by solid lines; flatter disks have lower fluxes at both wavelengths. Models with the same values of $H_p(R_0)$ and $\xi$ are connected by dashed lines. Models mod2b, mod2b,small and mod2b,large are all shown with filled squares. Models mod2b$^1$ and mod2b$^2$, characterized by different grain opacity, are shown by empty squares, models with larger grains have larger mm flux and smaller FIR fluxes. For easy comparison, symbols from Fig.~\ref{fig:fFIR_vs_fmm} are included in gray.}
\label{fig:colors}
\end{figure}

As in other studies of disk SEDs for TTS \citep[see, for example,][]{sic15}, we find that, for fixed stellar poperties, disks are colder when they are less flared or have higher mass. The midplane temperature in the outer disks is rather low, but never smaller than what is expected for the same grains heated only by the IRF ($\sim$7.5\,K). The (sub)-mm fluxes depend mostly on the disk mass, with a very weak dependence on the other parameters. The dependence on $L_\star$, in particular, is only weak. On the contrary, the FIR fluxes depend strongly on the temperature and, therefore, on $L_\star$, the disk shape, and also, although to a lesser degree, on the disk mass, as this also affects the temperature. In our grid, for the adopted opacity, the 100\,\mum\ optical depth in the vertical direction is $<$1 for $R$$>$2\,AU only for the lowest disk mass. However, as is always the case, a non-negligible portion of the emission may come from the optically thin, hotter surface layers \citep[see, e.g.,][]{liu15}. These trends are summarized in Fig.~\ref{fig:colors}, which plots the 100\,\mum\ {and 160\,\mum\ }fluxes as a function of the 1300\,\mum\ flux for all the models.
The figure shows also that both FIR and (sub)-mm fluxes do not depend strongly on \Rout\ nor on reasonable variations of the grain opacity. Another result worth noticing is that changes in the disk shape can more than compensate variations of $L_\star$, as shown by the fact that flat TTS disks lie in the same region of the diagram than flared BD disks.

\section{Discussion}
\subsection{Correlation between FIR and (sub)-mm fluxes}\label{sec:fFIR_vs_fmm}
We have shown in Sect.~\ref{sec:phot} (Fig.~\ref{fig:fFIR_vs_fmm}) that the observed FIR fluxes of our VLMO sample scale positively with the mm fluxes. A similar trend is also seen in Herbig Ae/Be stars \citep{pas16}. {How can this be interpreted in the light of our modeling result that FIR fluxes do not scale strongly with disk mass for a star of fixed luminosity?}

Radiation transfer models show that the (sub)-mm flux depends primarily on disk mass and only weakly on the other parameters. In contrast, FIR fluxes depend mostly on the temperature distribution, which itself is controlled by the stellar luminosity and disk shape. The dependence of the FIR fluxes on disk mass is weak, becoming only slightly stronger in more flared disks.
In particular, the dependence of FIR on disk mass is much weaker than the linear relation between $F_\mathrm{mm}$ and $M_\mathrm{d}$ as given by eq.~(\ref{eq:eq2}), which accordingly does not serve to correlate FIR and $F_\mathrm{mm}$ disk emission. Self-consistent hydrostatic equilibrium models \citep[e.g.,][]{sic15} show that, for fixed stellar parameters, disks of increasing mass are more flared and, therefore, have stronger FIR fluxes. However, even for fully flared disks the predicted correlation between the FIR flux and the (sub)-mm flux for a star of fixed luminosity is much weaker than linear.

\begin{figure}[tb]
\centering
\includegraphics[width=1\columnwidth]{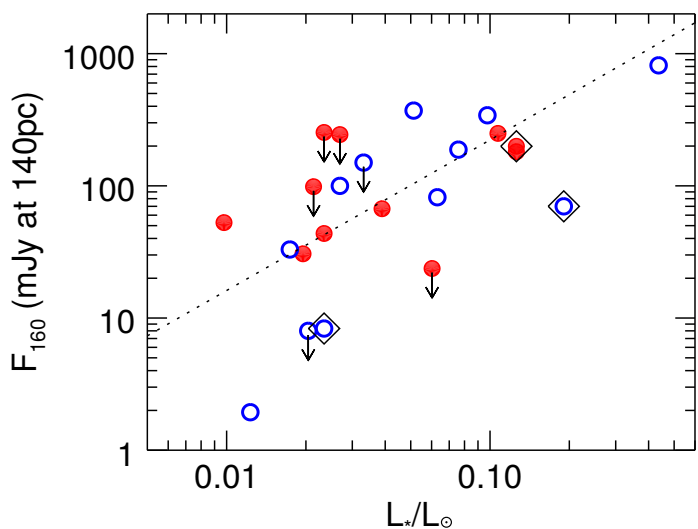}
\caption{\label{fig:fFIR_vs_Lstar} $F_{160}$ flux as a function of target luminosity. Symbols as in Fig.~\ref{fig:fFIR_vs_fmm}. Best linear fit to all values excluding upper limits and binaries shown as dashed line. Measured slopes $m_\mathrm{FIR}$ of best linear fits $\log(F_\mathrm{FIR})\propto m_\mathrm{FIR}\log(L_\star)$ in all FIR bands: $m_{70}$\,=\,1.30$\pm$0.25, $m_{100}$\,=\,1.09$\pm$0.39, and $m_{160}$\,=\,1.14$\pm$0.24.}
\end{figure}

It is noteworthy that optical depth apparently plays a subordinate role for the FIR--\Mdisk\ relation; despite the fact that the least massive disks are opticaly thin at most radii ($>$2\,AU at 100\,\mum) and that optically thin layers always contribute to the overall flux, even in regions where the vertical optical depth is $>$1, our models show no strong correlation of FIR with disk mass. We therefore provide an alternative explanation linking FIR emission directly to the total luminosity of the VLMOs.
We suggest that the apparent correlation between $F_\mathrm{FIR}$ and $F_\mathrm{mm}$ is in fact the result of two underlying correlations, namely the mass-luminosity relation for young VLMOs and a correlation of disk mass with stellar mass, paired with the observed strong correlation (Spearman's rank correlation parameters between 0.5 and 0.8 of all detections excluding binaries) of $F_\mathrm{FIR}$ and $L_\star$ (Fig.~\ref{fig:fFIR_vs_Lstar}).
What is observed in Fig.~\ref{fig:fFIR_vs_fmm} could then be traced back to a correlation $F_\mathrm{mm}$\,$\leftrightarrow$\,\Mdisk\ , according to eq.~\ref{eq:eq2}, a scaling between \Mdisk\ and $M_\star$ \citep[assumed to be linear, e.g.,][]{and13}, and a $M_\star$--$L_\star$ relation \citep[$L_\star$\,$\approx$\,$M_\star^{1.53}$ determined for 1\,Myr isochrones by][]{bar15} (the result is not sensitive to the particular $M$--$L$ correlation). The predicted slope $m$ of the correlation between $F_\mathrm{mm}$ and $L_\star$ of $m$\,=\,0.65\,$[\log(\mathrm{mJy})/\!\log(L_\odot)]$ is, however, shallower than what is seen in the observations ($m$\,=\,1.30$\pm$0.27\,$[\log(\mathrm{mJy})/\!\log(L_\odot)]$). The weak additional dependence between mm fluxes and other parameters (in particular, the temperature) or a stronger dependence between disk mass and stellar mass can make the trend slightly steeper and closer to the observed one. In fact, the most recent study of the \Mdisk--$M_\star$ relation \citep{ans16} suggests that the correlation is steeper than linear with a coefficient of 1.7--1.8 (for regions with ages of 1--2\,Myr). Their sample extends to masses of about 0.1\,\Msun. If this correlation continues into the substellar regime, it would bring our estimate above for the trend between FIR and $F_\mathrm{mm}$ in agreement with the observations.

\subsection{Spectral slopes in models and observations}\label{sec:flaring}
\begin{figure}[tb]
\centering
\includegraphics[width=1\columnwidth]{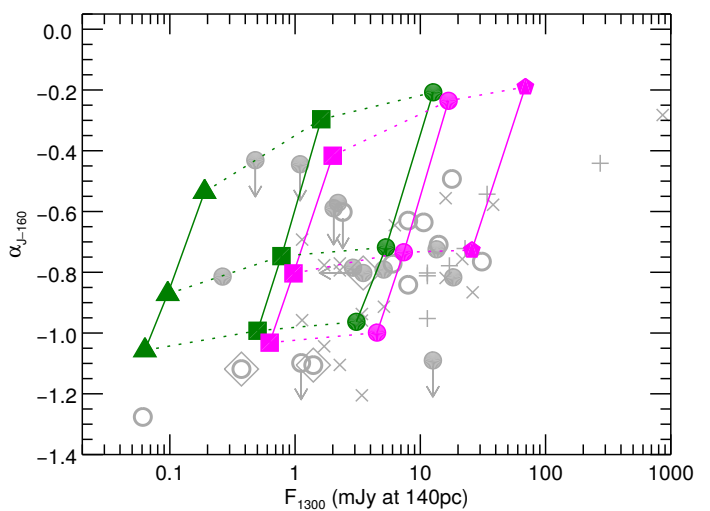}
\caption{\label{fig:IRslope_vs_mm_withmodels}
Model predictions of infrared slopes as a function of mm flux. Colors and symbols as in Fig.~\ref{fig:colors}. For reference, our measurements from the bottom panel of Fig.~\ref{fig:IRslope_vs_mm} were included in gray.}
\end{figure}

In Fig.~\ref{fig:IRslope_vs_mm_withmodels} we compare the spectral slopes derived from the observations (in gray) with the numbers produced by the models.
This is similar to the bottom panel of Fig.~\ref{fig:IRslope_vs_mm}, but with models overplotted. We show here the $J$$-$160 slopes because most literature objects do not have a 100\,\mum\ flux. Our results do not depend on that particular choice. As explained in Sect.~\ref{sec:models}, the shape of the disk is varied in the models by changing the flaring index and the scale height $H_0$, which are both listed in Table~\ref{tab:tab4}. The models cover three cases, from fully flared to very flat disk. 
The spectral slope depends strongly on the disk shape, but little on other parameters and, particularly, not on the disk mass and stellar luminosity.

From Fig.~\ref{fig:IRslope_vs_mm_withmodels} it is clear that the observed slopes are well reproduced by the models for moderately flared and flat disks. The fully flared disks, on the other hand, with a very high flaring index of 1.35 and large $H_0$, predict spectral slopes that are not seen in our sample. Since our sample contains objects that are bright in the FIR (i.e., they are detected with Herschel), it is unlikely that objects with higher spectral slopes exist but are missing. A comparison with literature samples in Sect.~\ref{sec:slopes} provides further confirmation. We conclude that the overwhelming majority of disks around very low-mass stars and brown dwarfs shows some flattening of the dust geometry compared to the hydrostatic case.

Our modeling results are in line with the findings of other groups. In \citet{sch06} it was already found that brown dwarf disks do not have the elevated flux levels expected for the hydrostatic case. Papers by \citet{har12}, \citet{alv13}, and \citet{liu15} fit SEDs with models calculated with a prescription analogous to ours for samples of mostly brown dwarfs (see Sect.~\ref{sec:obs} for more details on their samples). These authors all find that the flaring index ranges from 1.0 to 1.2 and the scaling factor $H_p(100\mathrm{\,AU})$ ranges from a few to 20, which again excludes the fully flared disks. \citet{liu15} also point out that the flaring index $\xi$ might be a function of spectral type (and thus stellar mass) in the brown dwarf regime with lower flaring indices for spectral types later than M8. Our sample does not extend to these late spectral types and therefore we cannot test this particular result.

\subsection{Disk masses from mm fluxes}\label{sec:Md_vs_fmm}
Before Herschel results were available, disk masses for brown dwarfs were mostly derived from single-band measurements at 0.85 or 1.3\,mm \citep[e.g.,][]{kle03,sch06,moh13}, using eq.~(\ref{eq:eq2}) with fixed temperature and opacity. At these wavelengths, the disks are optically thin and the emitted flux can be assumed to be proportional to the dust masses. For this calculation, it is assumed that dust grains emit as blackbodies; realistic values for opacity and dust temperature have to be adopted. This is the method we used in Sect.~\ref{sec:diskmasses} to estimate the disk masses in our sample. 

\begin{figure}[tb]
 \centering
  \includegraphics[width=\columnwidth]{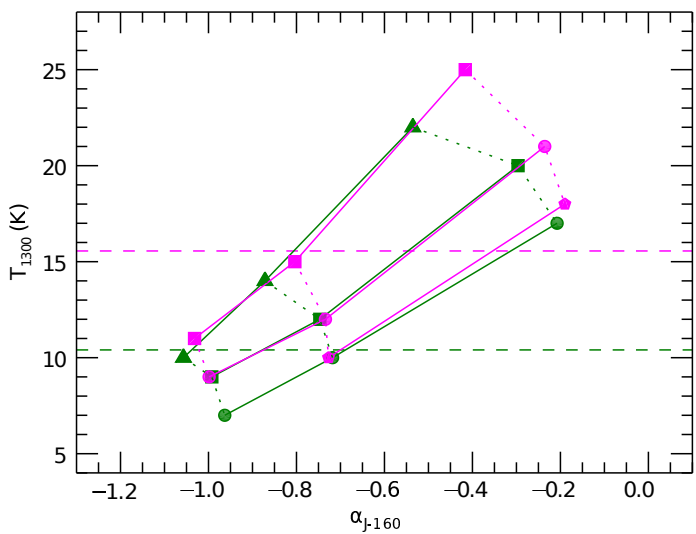}
 \caption{Values of the temperature needed to recover the model disk mass from the 1300\,\mum\ flux $T_{1300}$ vs. $\alpha_{J-160}$ for BD and TTS models. Colors and symbols as in Fig.~\ref{fig:colors}. $T_{1300}$ increases in more flared disks and decreases with the disk mass. TTS models, which have $L_\star$ that is five times larger than BD models, have higher $T_{1300}$ however, there is a large region of overlap. The dashed lines show the values of $T_{1300}$ according to the \citeauthor{and13} prescription ($T_\mathrm{A} = 25(L_\star/L_\odot)^{1/4}$\,K), green for the BD luminosity and magenta for TTS.}
\label{fig:temps}
\end{figure}

With the models presented in Sect.~\ref{sec:models} we can check the validity of the assumptions for the blackbody-based disk masses and evaluate uncertainties. In Fig.~\ref{fig:temps} we show the dust temperature required to reproduce the model disk mass with the blackbody-based prescription used in Sect.~\ref{sec:diskmasses} from the 1300\,\mum\ flux ($T_{1300}$) as function of the slope $\alpha_{J-160}$ for the BD and TTS models.
Very similar results are obtained for the 890\,\mum\ flux. The values of $T_{1300}$ reflect the overall temperature distribution in the disk; they are larger for more flared disks, as shown by the positive correlation of $T_{1300}$ with $\alpha_{J-160}$. The value $T_{1300}$ is lower for more massive disks {because stellar radiation does not penetrate the disk as deeply and the average temperature (at a given distance from the star) remains low}. For some models, $T_{1300}$ is very low, well below 10\,K. Only very flared, low-mass BD and TTS disks have $T_{1300}\gtrsim20$\,K. If we consider the observed range of $\alpha_{J-160}$, we can conclude that for most disks the appropriate values of $T_{1300}$ are in the range 10--15\,K. Higher values do not seem justified.

\begin{figure}[tb]
 \centering
 \includegraphics[width=\columnwidth]{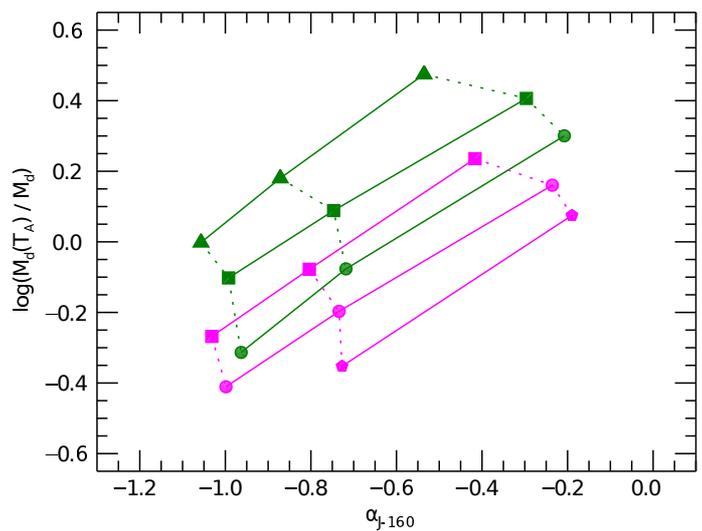}
 \caption{Ratio of the disk mass computed from $T_A$ over model disk mass as a function of $\alpha_{J-160}$. Colors and symbols as in Fig.~\ref{fig:colors}.}
\label{fig:BBdiskmass_vs_temp}
\end{figure}

In Fig.~\ref{fig:temps}, the horizontal lines indicate the temperatures expected for the BD and TTS luminosity adopted in our models, based on the scaling law derived by \citet{and13}, which we have also used in Sect.~\ref{sec:diskmasses}. The values are $\sim$10 and 15\,K, respectively. Most models in the observed range of $\alpha_{J-160}$ have $T_{1300}$ within this range and the adoption of their scaling law therefore appears justified. Fig.~\ref{fig:BBdiskmass_vs_temp} plots, as a function of $\alpha_{J-160}$, the difference between the values of \Mdisk\ obtained with $T_A$ from the \citeauthor{and13} prescription and the model disk mass.
The difference is typically less than a factor two, which is the maximum error that is made by adopting the blackbody assumption with the luminosity-scaled dust temperature.

However, it should be noted that many BD and TTS models have very similar values of $T_{1300}$, in spite of the factor 5 difference in $L_\star$. The \citeauthor{and13} $T_{1300}$ values, which scale only with $L_\star$, may introduce a spurious trend in the disk mass, especially if extended to very low $L_\star$, when $T_{1300}$ may be nominally lower than the value of $\sim$7\,K owing to heating by the IRF. A more accurate value of $T_{1300}$ can be obtained if $\alpha_{J-160}$ or any other FIR slope is known. 
In the case in which only FIR fluxes and no mm fluxes are available one requires an accurate stellar luminosity and a careful assessment of the temperature structure of the disk, since FIR fluxes depend strongly on temperature and disk shape and only weakly on disk mass (see Fig.~\ref{fig:colors}).

Very recently, \citet{van16}   proposed a new prescription to derive the disk mass from single wavelength (sub)-mm fluxes. They favor a scaling law with $L_\star$  that is flatter than the \citet{and13} law, namely $T_\mathrm{d}=22\,\mathrm{K}\,(L_\star/L_\odot)^{0.16}$, based on a grid of models with fixed shape ($H_0(100\mathrm{AU})=10$\,AU, $\xi=1.125$).  The difference between the two laws is larger at lower $L_\star$, becoming  a factor $\sim$1.6 for $L_\star=0.001\,L_\odot$.  The results obtained with the two scaling laws are compared to the disk mass derived by fitting the SED of eight very low luminosity ($L_\star \lesssim0.01\,L_\odot$) objects in the Upper Scorpius star-forming region. Masses derived with the \citeauthor{and13} scaling law are on average 3.5 times larger than the values {assumed in their radiative transfer model}, a discrepancy that is significantly reduced when using the new scaling law. 
It is possible that this is due in large part to the fact that disks in Upper Scorpius tend to be more settled and flatter than disks in younger star-forming regions such as Taurus \citep{sch07,sch12}. If so, the \citeauthor{van16} results confirm our conclusion that the disk shape is an important factor when deciding on the best value of $T_\mathrm{d}$ to adopt. However, the differences in disk masses are in general not very large.

We conclude that the major drawback in adopting a single scaling law to measure disk masses from (sub)-mm fluxes for a sample of objects is not so much in the error on individual objects but in the trend that this automatically introduces when comparing disks around stars of different luminosity and mass, such as those shown in Fig.~\ref{fig:disk_mass}.

\section{Summary}
We present far-infrared and (sub)-mm fluxes of 29 very low-mass stars and brown dwarfs with masses ranging from 0.03 to 0.2 solar masses. For 11 objects from this sample, we measured new FIR fluxes from Herschel PACS images. In addition, we compiled Herschel and (sub)-mm fluxes from the literature and rederived stellar parameters. The objects in this sample have detections in the FIR \emph{and} in the (sub)-mm as well as well-characterized stellar parameters. To interpret the trends seen in the observations, we model the SED for fiducial brown dwarf/star-disk systems using the radiation transfer code RADMC-3D. 

We show that VLMO disk masses can be robustly estimated from single wavelength (sub)-mm fluxes, and assuming blackbody emission, if the adopted dust temperature scales with luminosity as proposed by \citet[$T=25\,\mathrm{K}\,(L_\star/L_\odot)^{1/4}$]{and13}. This is the case at least for dust temperatures above $\sim$7\,K corresponding to the temperature of large grains in a typical interstellar radiation field. When choosing a temperature in this way, the error in the disk mass caused by assuming emission from a single temperature blackbody is at most a factor of two. The error is only larger for fully flared disks, which seem to be extremely rare. With additional information on the spectral slope in the FIR, these uncertainties can be further reduced. One should keep in mind that major uncertainties in the disk mass are introduced by the poor knowledge of dust opacity and gas-to-dust ratio.

From the (sub)-mm fluxes we estimate disk masses, adopting the aforementioned scaling law between dust temperature and stellar luminosity. The disk masses in our sample range from 10$^{-4}$ to 10$^{-2}$ solar masses. More than half of our sample has a disk mass above 10$^{-3}$ solar masses (corresponding to about 1 Jupiter mass). While our sample is biased toward high disk masses, this shows that a fraction of brown dwarfs has substantial disk masses, as already found by other authors \citep{sch06,bou08,ric14}. The disk mass is critical for the outcome of planet-forming processes; simulations by \citet{pay07} indicate that Jupiter-mass disks around brown dwarfs have the potential to form Earth-mass planets. As found by other authors, the disk masses in our sample are in the range of 1\% of the object mass, which is similar to what is found in more massive stars.

A surprising find of our study is that the observed FIR fluxes scale with the (sub)-mm fluxes. This is not expected from disk models because FIR fluxes mostly depend on the temperature structure, whereas (sub)-mm fluxes mostly scale with disk mass. We show that this trend can be qualitatively explained as a result of three interlinked correlations: the strong link between FIR fluxes and stellar luminosity, the stellar mass-luminosity relation, and the scaling of disk mass with stellar mass. 

The comparison of the NIR-FIR spectral slopes ($\alpha_{J-\mathrm{FIR}}$) with the models clearly shows that brown dwarf disks are not fully flared, which is indirect evidence {for the settling of large grains to the disk midplane}. There is no evidence for a mass-dependence in the NIR-FIR spectral slope, neither in our sample nor when comparing with values measured for more massive T Tauri stars. We note that this does not necessarily mean that the physical processes that determine the disk shape occur in the same way in brown dwarfs and stars \citep[see the discussions in][]{szu10,pas09,mul12}.

\begin{acknowledgements}
  We thank the unknown referee for a thoughtful review.
  Thanks to Aurora Sicilia-Aguilar and Michael Meyer for instructive comments on the manuscript. 
  AN would like to acknowledge funding from Science Foundation Ireland (Grant 13/ERC/I2907).
  AS acknowledges support from STFC grant ST/M001296/1. This work was partly supported by the Italian Ministero dell'Istruzione, Universit{\`a} e Ricerca through the grant Progetti Premiali 2012 – iALMA (CUP C52I13000140001). 
  Our research was also supported by NSERC grants to RJ. 
  Research visits by SD and AN to the University of St Andrews and AS to the Dublin Institute for Advanced Studies have greatly facilitated the work on this paper. We are grateful for the hospitality and the generous technical/financial support by these two institutions. 
  This research has made use of the SIMBAD database and the VizieR catalog access tool, operated at CDS, Strasbourg, France.
\end{acknowledgements}

\appendix
\section{Spectral energy distributions}\label{sec:appA}
\begin{figure*}[b]
\includegraphics[width=\textwidth]{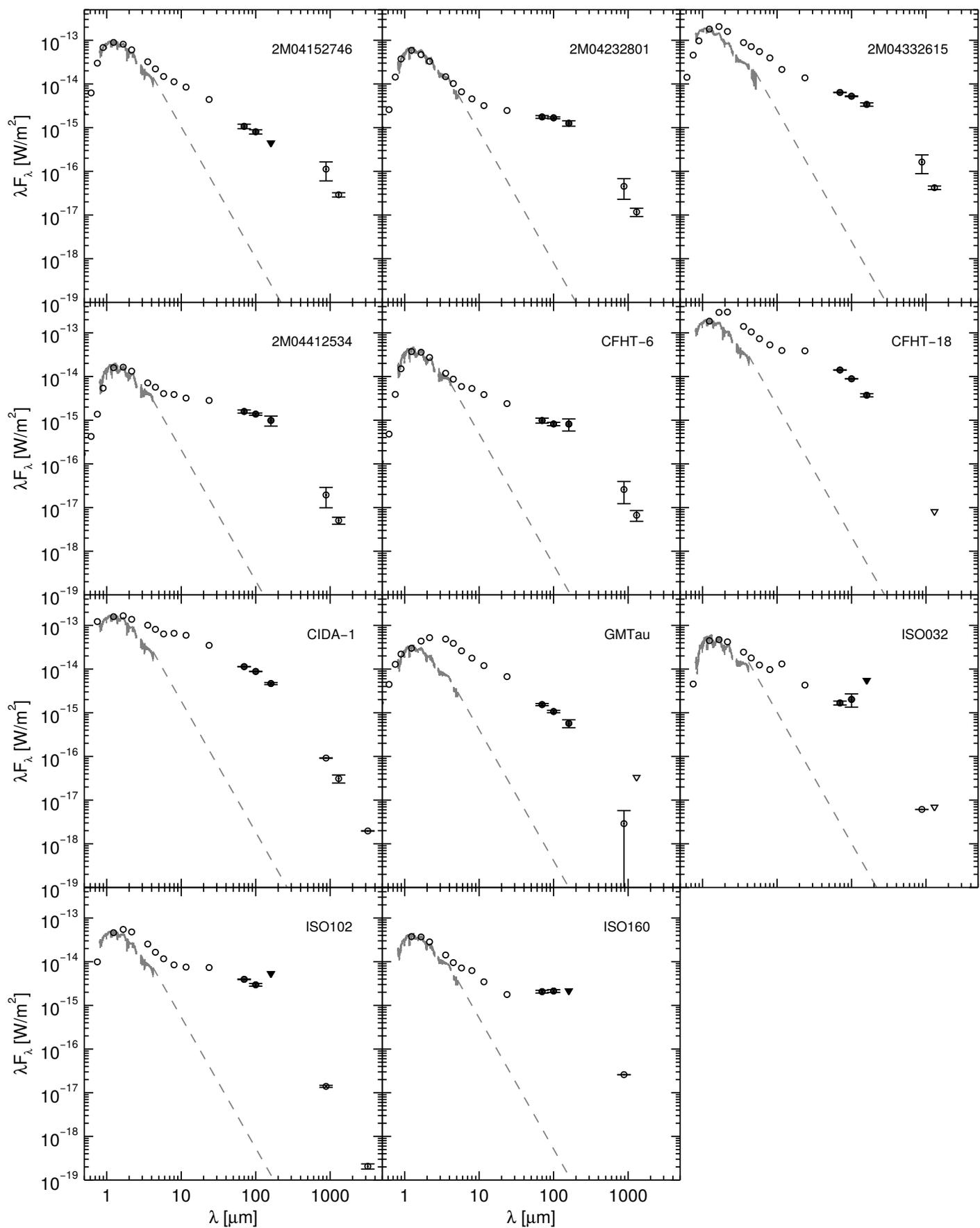}
\caption{\label{fig:SEDs}Dereddened SEDs of our core sample. The filled symbols show data at Herschel wavelengths, the open symbols are literature values at other wavelengths. Upper limits are shown with triangles. A stellar SED \citep[from the IRTF Spectral Library;][]{cus05,ray09}, normalized at J band, is shown for comparison.}
\end{figure*}

\begin{figure*}[b]
\includegraphics[width=\textwidth]{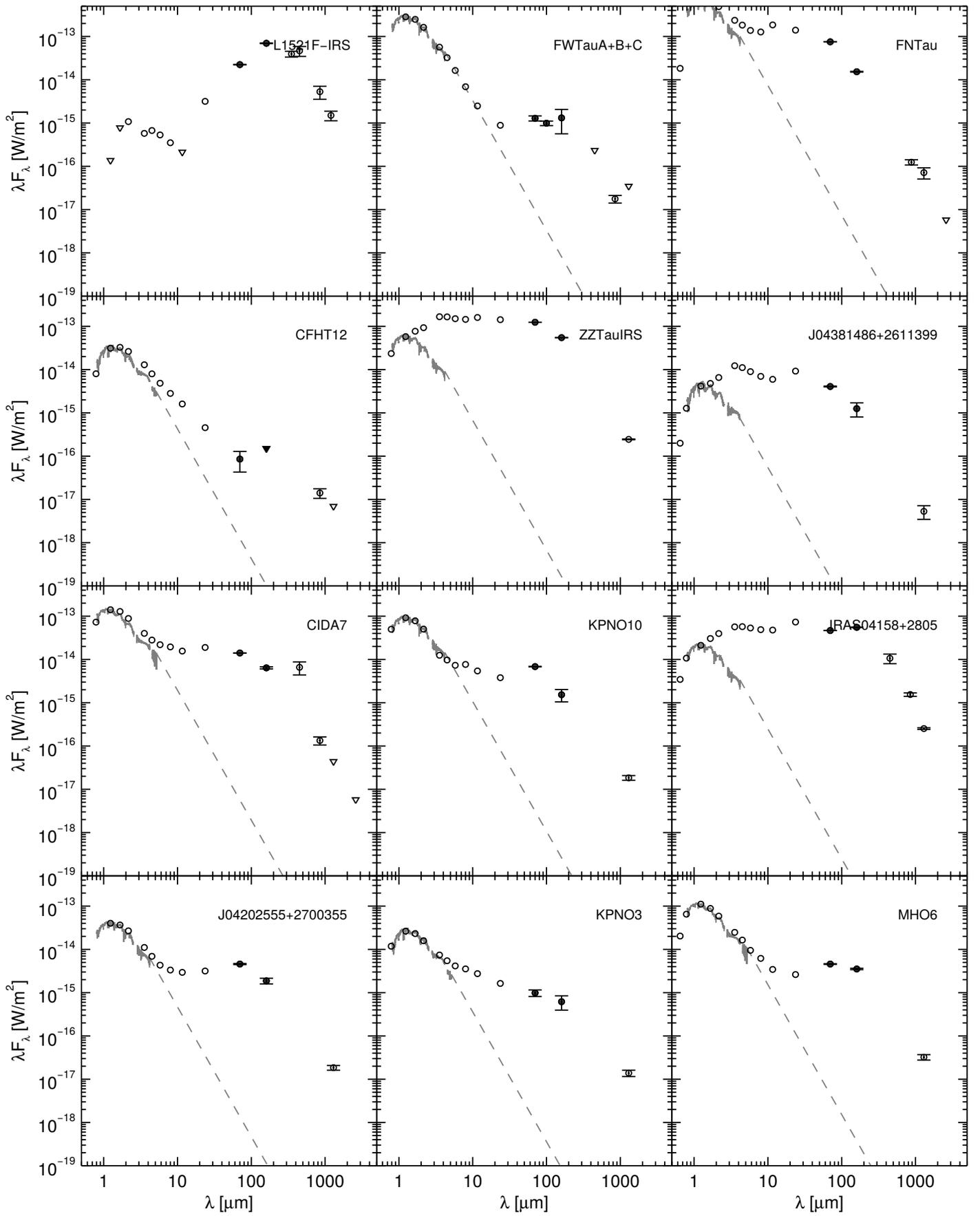}
\caption{\label{fig:SEDsLit}SEDs of our literature sample. Symbols and lines as in Fig.~\ref{fig:SEDs}.}
\end{figure*}

See Figs.~\ref{fig:SEDs} and \ref{fig:SEDsLit} for spectral energy distributions of all targets presented in this paper.

\addtocounter{figure}{-1}
\begin{figure*}[b]
\includegraphics[width=\textwidth]{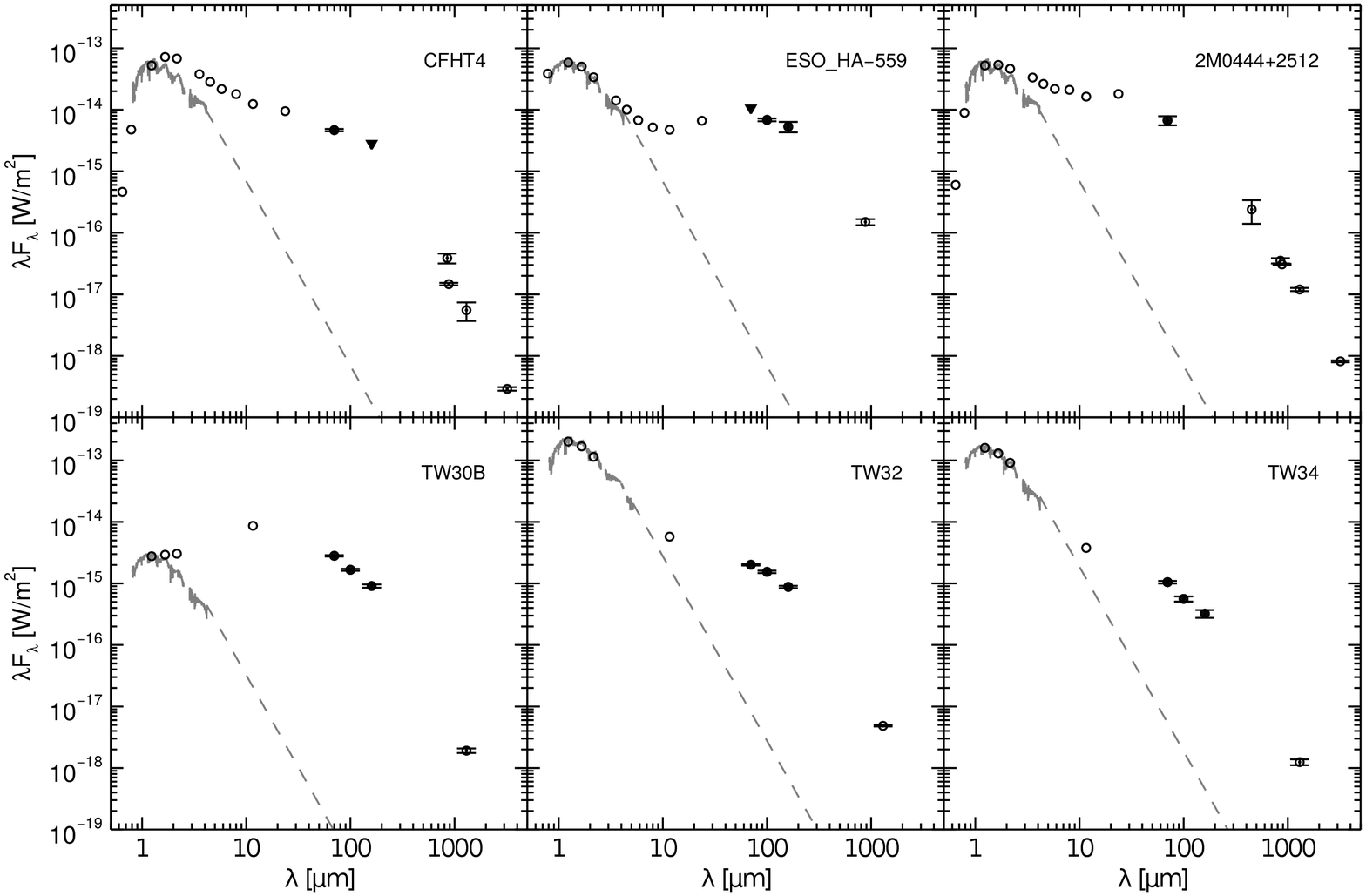}
\caption{\emph{Fig.~\ref{fig:SEDsLit} ctd...}}
\end{figure*}

\end{document}